\documentclass[conference]{IEEEtran}
\IEEEoverridecommandlockouts
\usepackage{cite}
\usepackage{amsmath,amssymb,amsfonts}
\usepackage{algorithmic}
\usepackage{graphicx}
\usepackage{textcomp}
\usepackage{comment}
\usepackage{xcolor}
\def\BibTeX{{\rm B\kern-.05em{\sc i\kern-.025em b}\kern-.08em
    T\kern-.1667em\lower.7ex\hbox{E}\kern-.125emX}}

\newenvironment{psmallmatrix}
  {\left(\begin{smallmatrix}}
  {\end{smallmatrix}\right)}

\usepackage{qcircuit}
\usepackage{graphicx}
\usepackage{subcaption}
\usepackage{float}

\begin{document}

\title{A Hybrid Quantum-Classical Framework for Utility-Scale Edge Detection of Real-World Medical and Geospatial Data
%A Hybrid Quantum-Classical Decomposition Framework for Utility-Scale Edge Detection \\
%Towards a Utility-Scale Quantum Edge Detection for Real-World Medical Image Data\\
{\footnotesize 
%\textsuperscript{*}Note: Sub-titles are not captured in Xplore and should not be used}
\thanks{This research was sponsored by the Richard T. Cheng Endowment, Old Dominion University.}
}
}

\author{\IEEEauthorblockN{Emmanuel Billias}
\IEEEauthorblockA{\textit{Center for Real-Time Computing} \\
\textit{Computer Science Department}\\
\textit{Old Dominion University}\\
Norfolk, VA \\
ebill007@odu.edu}
\and
\IEEEauthorblockN{David Calano}
\IEEEauthorblockA{\textit{Web Science \& Digital Libraries} \\
\textit{Computer Science Department}\\
\textit{Old Dominion University}\\
Norfolk, VA \\
dcala005@odu.edu}
\and
\IEEEauthorblockN{
Nikos Chrisochoides}
\IEEEauthorblockA{\textit{Center for Real-Time Computing} \\
\textit{Computer Science and Physics Departments}\\
\textit{Old Dominion University}\\
Norfolk, VA \\
nikos@cs.odu.edu}
}

\maketitle

\begin{abstract} 
%% Version 3 (AI generated from Version 2) for clarity and English Prof.

We present a \textbf{hybrid quantum-classical framework} designed to achieve \textbf{utility-scale} performance for Quantum Hadamard Edge Detection (QHED) on Noisy Intermediate-Scale Quantum (NISQ) devices. The framework utilizes a \textbf{Two-Level Decomposition} strategy: (1) \textbf{Problem-Level Decomposition (PLD)}, which partitions high-resolution \textbf{real-world data} into buffered sub-images, and (2) \textbf{Circuit-Level Decomposition (CLD)}, which employs circuit-cutting to reduce complexity for near-term hardware. This approach, combined with a depth-efficient \textbf{$QHED^M$} decrement gate, achieves a \textbf{62\% reduction in circuit depth} and \textbf{93\% fewer two-qubit operations}. We demonstrate the framework's \textbf{domain-agnostic utility} by processing \textbf{real-world data} from Medical Image Computing (MIC) and Geospatial Information Systems (GIS). Crucially, we investigate the \textbf{resilience crossover point} by comparing a $[\![3,1,1]\!]$ bit-flip repetition code against passive \textbf{Quantum Error Mitigation (QEM)}. Our results indicate that for current coherence times, passive mitigation offers a superior utility advantage by bypassing the gate-overhead penalties inherent in active encoding, recovering nearly \textbf{99\% of the ideal signal}.

\end{abstract}

\begin{IEEEkeywords}
Quantum Computing, Quantum Edge Detection, Algorithm, Circuit Cutting, Circuit Knitting, Problem Decomposition, Quantum Utility-Scale, Medical Image Computing
\end{IEEEkeywords}

\section{Introduction}

Quantum image processing has significant potential to revolutionize medical imaging, particularly in applications such as image-guided neurosurgery (IGNS) for brain cancer \cite{Chrisochoides_Liu_Drakopoulos_Kot_Foteinos_Tsolakis_Billias_Clatz_Ayache_Fedorov_et_al._2023a}. This involves deformable registration, a minimization problem, of preoperative and intraoperative MRI data in real-time, where registration accuracy and speed are critical. Edge detection, a fundamental image analysis task, is crucial in identifying critical anatomical structures\cite{Liu_Kot_Drakopoulos_Yao_Fedorov_Enquobahrie_Clatz_Chrisochoides_2014}. However, achieving accurate and efficient edge detection on Noisy Intermediate-Scale Quantum (NISQ) hardware presents substantial challenges. 
Due to large volumes of data for high-resolution images, effective processing methods continually require refinement\cite{ma}. On the other hand, naive quantum approaches often suffer from limitations related to noise sensitivity and qubit resource constraints, leading to results that are impractical for real-world applications\cite{Le2011}.

This paper addresses these challenges by proposing a novel hybrid quantum-classical methodology based on a 2-level decomposition strategy. We introduce techniques to optimize image encoding for quantum circuits at the data level. Specifically, we leverage buffer pixels to mitigate artifacts introduced by image decomposition, a necessary step for mapping realistic images onto scalable NISQ architectures. This approach enhances the spatial locality of data, improving subsequent quantum processing output. 

At the circuit level, we focus on enhancing the fidelity and efficiency of the Quantum Hadamard Edge Detection (QHED) algorithm\cite{yao}. First, we restructure the QHED circuit using a more depth-efficient decrement permutation gate. We then employ circuit-cutting techniques \cite{tang2025tensorqcscalabledistributedquantum}\cite{10.1145/3445814.3446758}\cite{10374226}\cite{10821102}\cite{10488877} to reduce the depth of quantum circuits and minimize the impact of noise, leading to significant improvements in fidelity and a substantial reduction in two-qubit operations.

The Distributed Noisy Intermediate-Scale Quantum (D-NISQ) methodology\cite{acampora_di} partitions the computational problem into multiple subdomains, each evaluated independently. Due to the potentially large number of subcircuits resulting from image decomposition, we implement a proof-of-concept D-NISQ simulation on a high-performance computing (HPC) cluster\cite{elsharkawy2023integrationquantumacceleratorshigh}\cite{10.1145/3674151}. Ongoing research into multinode superconducting quantum computer architectures\cite{BARRAL2025100747}, as well as efforts to integrate quantum computing into existing HPC ecosystems\cite{BECK202411}, demonstrate the increasing convergence of classical and quantum computing paradigms. These architectures share several key characteristics: a centralized scheduler coordinates the distribution, execution, and aggregation of computational subdomains, while classical and quantum resources are orchestrated to optimize overall system utilization. Quantum operations are offloaded to quantum processing units (QPUs) or logical partitions thereof, with the resulting data collected and integrated into the broader computational workflow.

\begin{comment}
Furthermore, we explore the feasibility of processing raw k-space Magnetic Resonance Imaging (MRI) data, preprocessed with the Inverse Quantum Fourier Transform (IQFT), using QHED under ideal conditions. The QFT is utilized for many notable quantum algorithms such as SShor'salgorithm ~\cite{doi:10.1137/S0036144598347011}, the quantum phase estimation algorithm ~\cite{kitaev1995quantum}, and the Abelian hidden subgroup problem ~\cite{ettinger1999quantum}. The optimized DFT algorithm offers a computational complexity of \(O(N^{2})\) where the QFT circuit reduces to \(O(NlogN)\) ~\cite{9198106}~\cite{892139}, making it a highly desirable algorithm to leverage. Combining IQFT and QHED with 2-level decomposition is meant to demonstrate the potential for integrating quantum image processing into clinical workflows where accurate real-time analysis is critical~\cite{ARCHIP2007609}. 
%such as Deep Brain Stimulation (DBS) \cite{Lozano_Lipsman_Bergman_Brown_Chabardes_Chang_Matthews_McIntyre_Schlaepfer_Schulder_et_al._2019}. 
This research aims to bridge the gap between non-rigid registration (NRR) using nested expectation maximization (NEM) \cite{Liu2014} with practical applications on NISQ devices \cite{Miyahara_2020}. By addressing the limitations of classical and naive quantum approaches, we seek to improve the accuracy and efficiency of edge detection, particularly for medical imaging tasks\cite{Liu2014}.
\end{comment}

As a final step, we demonstrate the benefits of the proposed QHED circuit in processing raw k-space magnetic resonance imaging (MRI) data, preprocessed using the inverse quantum Fourier transform (IQFT) instead of classical DFT, with the hope of saving conversion costs from classical data in the future. Given that the optimized DFT algorithm has a computational complexity of \(O(N^{2})\), while the QFT circuit reduces this to \(O(N \ log N)\) for a given number of \(n=log_{2}(N)\) qubits\cite{9198106}\cite{892139}, it is a highly desirable next step to test the capabilities of our approach and demonstrate the potential for integrating quantum image processing into clinical workflows where accurate real-time analysis is critical~\cite{ARCHIP2007609}.

The remainder of this paper is structured as follows: Section 2 provides background on QHED circuit and circuit cutting. Section 3 details our improvement on the QHED circuit, data decomposition methodology, circuit cutting, and two-level decomposition as a whole. Section 4 presents the experimental setup and results for the circuit analysis and results on real MRI images. Section 5 concludes the paper and outlines future work.

\noindent \newline
{\bf Contributions}
The contributions of this paper are as follows: 

\begin{itemize}
\item modification of the decrement permutation of the QHED circuit and introduction of optimizations to maintain over 96\% circuit fidelity (as opposed to 89\% for Yao's et al. method \cite{yao}) using realistic noise models from IBM for 5-qubit data input sizes.

\item introduction of a new two-level image and circuit decomposition strategy, resulting in a reduction of the sub-circuit depth and two-qubit operations by approximately 62\% and 93\%  (compared to Yao's et al. method), respectively.

\item extention of work by Yao et al. which utilized smaller, synthetic binary images, by successfully applying an image-level decomposition to both 2D (1024 x 1024 pixels) and 3D brain Magnetic Resonance (MR) images (256 x 256 x 130 voxels). This demonstrates the feasibility and potential of applying the proposed approach for complex real-world medical imaging data, a significant step beyond previous studies focused on simplified benchmarks.

%\item We evaluated the new two-level decomposition of the decrement permutation circuit with realistic images (as opposed to toy synthetic benchmarks) on today's noisy quantum processor units (QPU), using 2D (1024 x 1024 pixels) and 3D brain MR images (256 x 256 x 130 voxels) which extends the work of Yao et al. \cite{yao} from smaller, binary images to real sized MRI data.

%% (Nikos) \item key lesson learned (see Figure \ref{fig99}) is that while reducing the width of the proposed quantum circuit leads to a significant decoupling of the correlation between the number of CX gates and the circuit's fidelity, this benefit highlights its practical utility in quantum circuit design, however it comes at the cost of classical overhead.

%\item From our evaluation, we derive the key lesson that circuit cutting can enable restructured architectures which help decouple fidelity from CNOT gate count, highlighting its practical utility in quantum circuit design.

%(2) we use buffer pixels to eliminate image artifacts introduced by the image decomposition, which is required to utilize scalable, distributed NISQ hardware; 

%\item We present preliminary results using real-size MRI images (as opposed to toy synthetic benchmarks): 2D (1024x1024 pixels) and 3D brain MR Image (256x256x130 voxels).
%a dynamically scalable multi-node, multi-core distribution for the simulated QHED algorithm the on a both a 2D (1024x1024 pixels) and 3D brain MR Image (256x256x64 voxels).
\end{itemize}
\section{Background}

\subsection{Quantum Hadamard Edge Detection}

\begin{figure}[h]
    \centering
    \includegraphics[width = 0.45\textwidth]{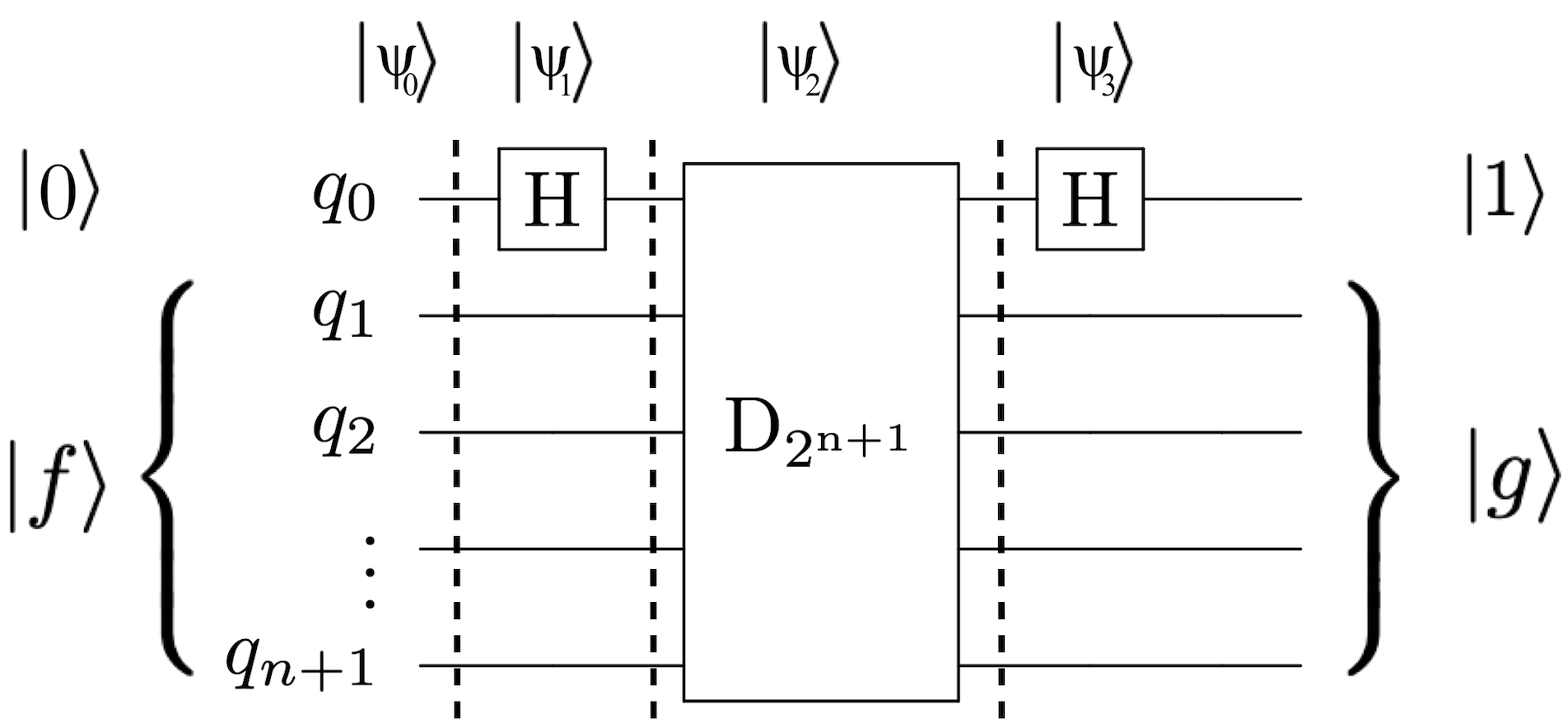}
    \caption{The QHED circuit with an ancillary qubit. We follow the convention that \(q_{0}\) is the LSB. The \(D_{2^{n+1}}\) gate applies an amplitude permutation, functioning as a decrement operation on the input state vector.}
    \label{fig1}
\end{figure}

\begin{figure*}
    \centering
    \includegraphics[width=0.95\linewidth]{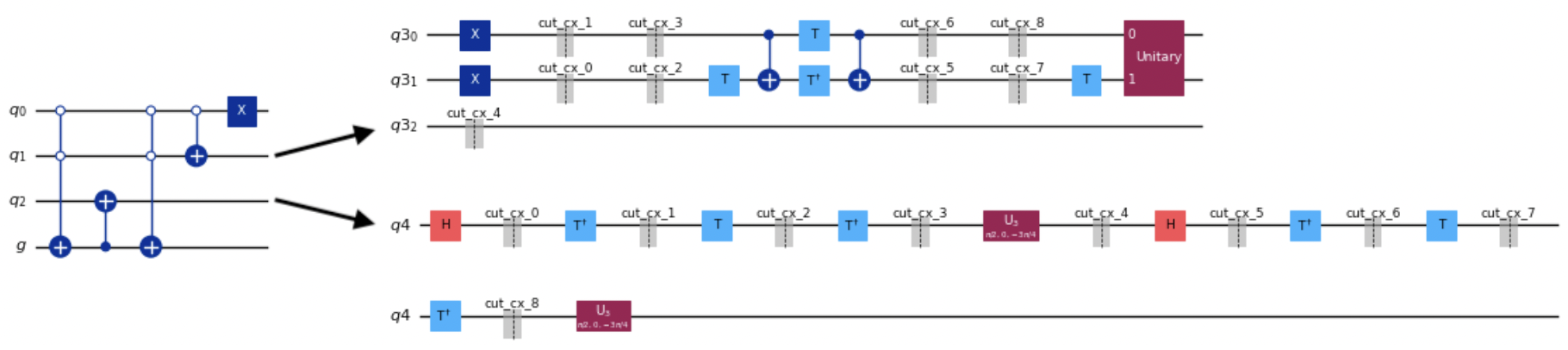}
    \caption{Example of circuit cutting: The modified decrement permutation (left) is partitioned into smaller subcircuits (right) generated by Qiskit circuit cutting, enabling execution on NISQ hardware for increased fidelity. The reconstructed results are obtained by post-processing measurements from the subcircuits.}
    \label{fig32}
\end{figure*}

Shown in Fig. \ref{fig1} is the full boundary QHED circuit which can be performed in \(O(1)\) time\cite{yao}. The 2D image is decomposed to 1D via row-major ordering for the algorithm to detect horizontal edges and column-major ordering for vertical edges. The state \(|\psi_{0}\rangle\) represents a quantum input state with a redundant \(|0\rangle\) least significant qubit (LSB). \(2^{n}\) bits of classical data are encoded into \(n\) number of qubits labeled \(q_{1}\) to \(q_{n+1}\) using the concept of amplitude encoding\cite{doi:10.1137/S0036144598347011}\cite{10.48550}. An input vector \(\Vec{f} = (f_{0}, f_{1},...,f_{2^{n}-1})\) is transformed into the normalized quantum state \( \left|f\right\rangle = (c_{0}, c_{1}, ..., c_{2^{n}-1})^{T} \) shown in  \eqref{eq:1}. This image is coupled with a redundant qubit in the least significant position initialized to state \(|0\rangle\), producing the state shown in  \eqref{eq55}.

\begin{comment}
\begin{figure}
    \centering
    \begin{subfigure}[b]{0.25\textwidth}
         \centering
         \includegraphics[width=\textwidth]{img/amp_img.png}
     \end{subfigure}
     \begin{subfigure}[b]{0.45\textwidth}
         \centering
         \includegraphics[width=\textwidth]{img/amp_circ.png}
     \end{subfigure}
    \caption{The image (top) can be normalized and encoded into a quantum circuit (bottom).}
\end{figure}
\end{comment}

\begin{equation}\label{eq:1}
\left|f\right\rangle=
\sum_{i=0}^{2^{n} - 1} c_{i}\left|i\right\rangle = \frac{f_{i}}{\sqrt{\sum_{k=0}^{2^{n}-1}f_{k}^{2}}}
\end{equation}

Applying a Hadamard gate to the LSB will produce the well-known \(\left|+\right\rangle\) state on the prepared vector. The redundant image state, \(|\psi_{1}\rangle\), is shown in  \eqref{eq66}. The Decrement unitary, shown in  \eqref{eq:3}, yields the new redundant state, \(|\psi_{2}\rangle\), shown in  \eqref{eq7}.

\begin{equation} \label{eq:3}
D_{2^{n+1}} =
    \begin{bmatrix}
    0 & 1 & 0 & 0 & \dots & 0 & 0\\
    0 & 0 & 1 & 0 & \dots & 0 & 0\\
    0 & 0 & 0 & 1 & \dots & 0 & 0\\
    \vdots & \vdots & \vdots & \vdots & \ddots & \vdots & \vdots\\
    0 & 0 & 0 & 0 & \dots & 0 & 1\\
    1 & 0 & 0 & 0 & \dots & 0 & 0
    \end{bmatrix}
\end{equation}

The final permutation is described as a Hadamard gate applied to the LSB after the decrement permutation. The resultant unitary, shown in  \eqref{eq:4}, can be considered to take the form of \(I\otimes I...\otimes H\).

\begin{equation} \label{eq:4}
    I_{2^{n}} \otimes H = 
\frac{1}{\sqrt{2}}
\begin{bmatrix}
1 &  1 & 0 &  0 & \cdots & 0 & 0\\
1 & -1 & 0 &  0 & \cdots & 0 & 0\\
0 &  0 & 1 &  1 & \cdots & 0 & 0\\
0 &  0 & 1 & -1 & \cdots & 0 & 0\\
\vdots &\vdots &\vdots &\vdots & \ddots & \vdots & \vdots\\
0 & 0 & 0 & 0 & \hdots & 1 & 1\\
0 & 0 & 0 & 0 & \hdots & 1 & -1\\
\end{bmatrix}
\end{equation}

This gate produces the final processed image state, \(|\psi_{3}\rangle\), shown in  \eqref{eq8}. In the case of edge detection as a subroutine, we only care for the odd-indexed basis states which occur when the LSB, \(q_{0}\), is in the \(\left|1\right\rangle\) state. The odd-indexed states result in the encoded differences of values between adjacent pixels, a form of edge detection in the case of image processing.

\begin{equation}\label{eq55}
    |\psi_{0}\rangle = |f\rangle \otimes |0\rangle
\end{equation}

\begin{equation}\label{eq66}
    |\psi_{1}\rangle = |f\rangle \otimes |+\rangle = \frac{1}{\sqrt{2}} \begin{bmatrix} c_{0} \\ c_{0} \\ c_{1} \\ c_{1} \\ \vdots
    \end{bmatrix}
\end{equation}

\begin{equation}\label{eq7}
    |\psi_{2}\rangle = D_{2^{n+1}}|\psi_{1}\rangle = \frac{1}{\sqrt{2}} \begin{bmatrix} c_{0} \\ c_{1} \\ c_{1} \\ \vdots \\ c_{0}
    \end{bmatrix}
\end{equation}

\begin{equation}\label{eq8}
    |\psi_{3}\rangle = (I_{2^{n}} \otimes H)|\psi_{3}\rangle = \frac{1}{2} \begin{bmatrix} c_{0} + c_{1} \\ c_{0} - c_{1} \\ c_{1} + c_{2} \\ \vdots \\ c_{2^{n} - 1} - c_{0}
    \end{bmatrix}
\end{equation}

\subsection{Circuit Cutting}

Circuit cutting is a method for decomposing larger circuits into smaller subcircuits, with the trade-off of additional sampling overhead\cite{tang2025tensorqcscalabledistributedquantum}\cite{10488877}\cite{qiskit-addon-cutting}. Circuit width, depth, and swaps can be decreased to execute the instruction set architecture (ISA) on hardware to improve performance. In our case, each of the \(P\)  encoded subcircuits resulting from the data-level decomposition is decomposed farther into a \(Q\) number of subcircuit resulting from the Circuit-Level Decomposition demonstrated in Fig. \ref{fig32}. The results of these smaller circuits can be recombined to yield results equivalent to those of the original circuit, albeit with ideally higher fidelity.

\section{Proposed NISQ Enhancements}

Quantum fidelity is a mathematical expression of one quantum state passing as another. In our case, we describe how a noisy QHED output compares to the ideal. Jozsa gives an account for Ulhmann's transition probability\cite{Uhlmann1976TheP} in the restricted finite dimensional space\cite{Jozsa_fidelity} shown in  \eqref{eq:5}. This describes fidelity between two quantum channels (quantum circuits) given their resultant density operators, also referred to as density matrices. Shown in  \eqref{eq:6}, each pure output state \(|\psi_{j}\rangle\) is sampled from a noisy or ideal circuit. Given an ensemble of output states, we can describe that each pure state has some output probability \(p_{j}\). Given both a noisy and ideal ensemble via sampling, the density matrices can be constructed and fidelity calculated.

\begin{equation} \label{eq:5}
    F(\rho, \sigma) = \left(tr\sqrt{\sqrt{\rho}\sigma\sqrt{\rho}}\right)^{2}
\end{equation}

\begin{equation} \label{eq:6}
    \rho = \sum_{j}p_{j}|\psi_{j}\rangle\langle\psi_{j}|
\end{equation}

Amplitude-encoded images produce exponential circuit depth regarding the input vector size\cite{araujo}. Furthermore, the decrement permutation required to make the final image state of size \(n+1\) qubits with the same number of \(n+1\) qubits is comprised of a series of multi-controlled NOT (MCX) gates\cite{li_yang_torres_zheng_wang_2013}. During transpilation, MCX gates are mapped as a series of CNOT and rotational gates. We can use several techniques to improve the overall fidelity of our circuit output. First, we modify the QHED circuit using ancillary qubits to produce a linear number of CX operations. We then use image decomposition, which scales the problem into \(P\) subdomains to save exponential circuit depth from the amplitude encoding. Finally, a series of circuit optimizations that will further enhance and cut each subdomain into \(Q\) number of circuits to further increase fidelity.

\subsection{Decrement Gate Alteration}

\begin{figure}[h]
     \centering
     \begin{subfigure}[b]{0.35\textwidth}
         \centering
         \includegraphics[width=\textwidth]{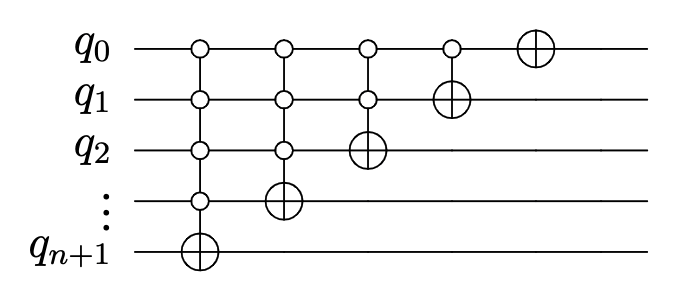}
     \end{subfigure}
     \hfill
     \begin{subfigure}[b]{0.45\textwidth}
         \centering
         \includegraphics[width=\textwidth]{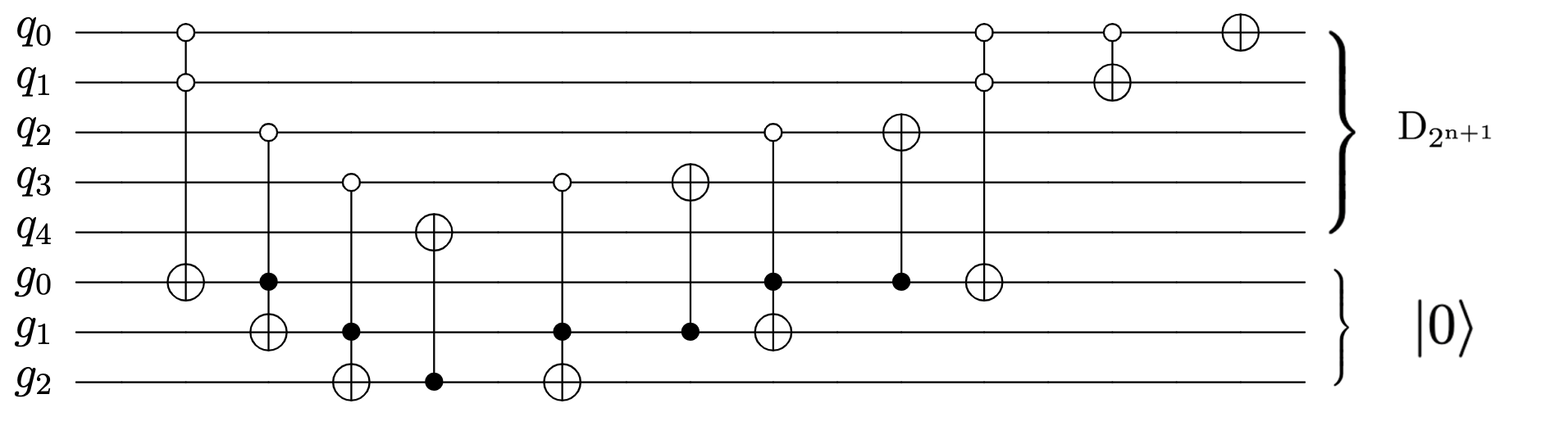}
     \end{subfigure}
     \caption{The decrement gate used in the QHED circuit (top) results in an exponential relationship between circuit depth and width. In contrast, the alternative decrement circuit (bottom) consists solely of CX and Toffoli gates, reducing to a linear number of CX gates after transpilation.}
    \label{fig5}
\end{figure}

\begin{figure}
    \centering
    \includegraphics[width=0.45\textwidth]{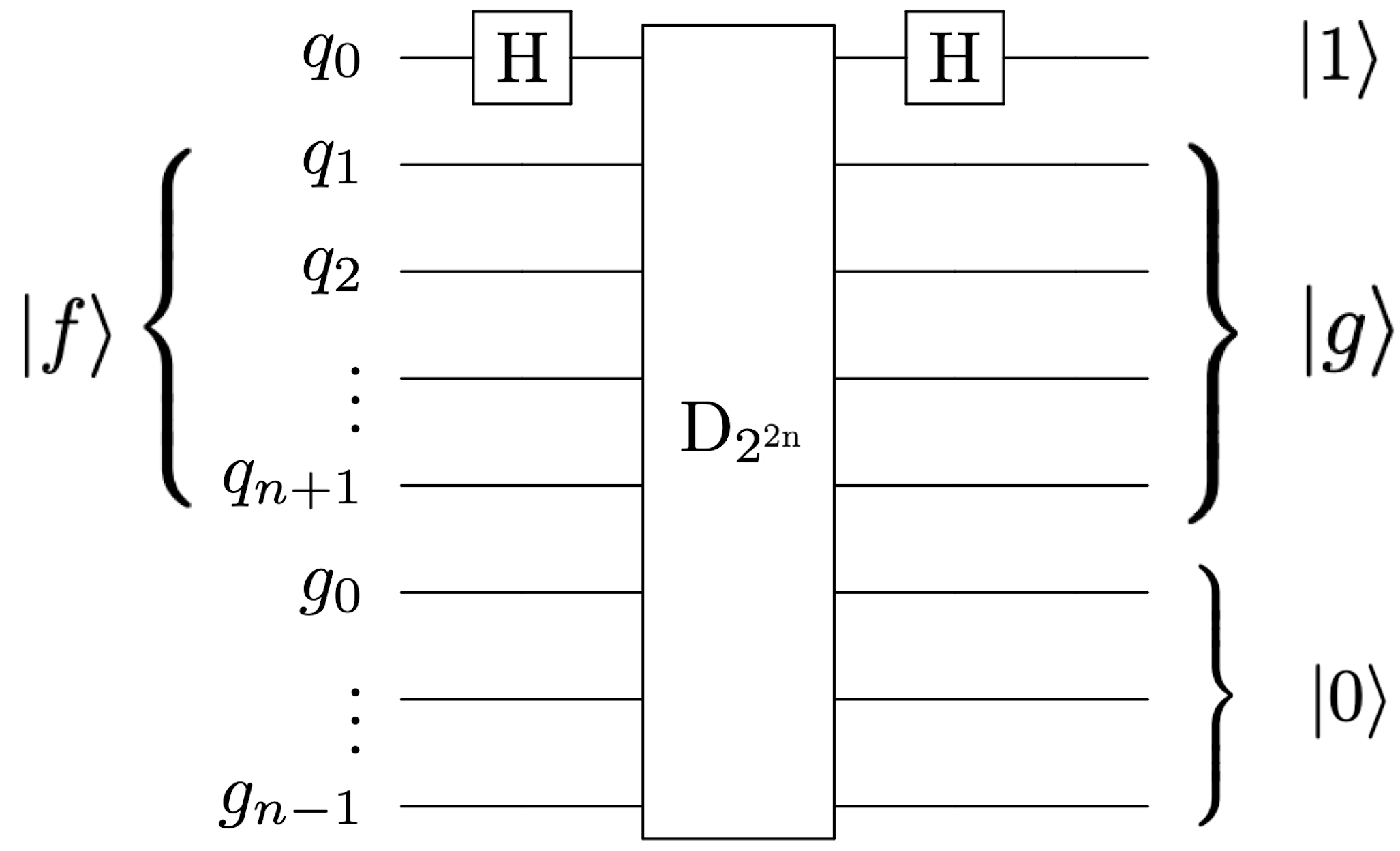}
    \caption{Our suggested QHED modification for quality enhancement on NISQ era hardware, utilizing ancillary qubits.}
    \label{fig6}
\end{figure}

A decrement gate can be explicitly constructed using an ascending series of multi-controlled NOT (MCX) gates \cite{li_yang_torres_zheng_wang_2013}. Since MCX gates must be further decomposed into a sequence of CNOT gates for execution on real hardware via the transpiler, an alternative approach leverages \(n - 2\) additionally ancillary qubits to implement the decrement permutation using only CX and Toffoli gates \cite{gidney}, as shown in Fig. \ref{fig5} generalized by \eqref{eq89}. This allows us to construct our modified QHED circuit, Fig. \ref{fig6}, referred to in the paper as QHED\textsuperscript{M}.

\begin{equation}\label{eq89}
    D_{2^{2n-2}}(|\psi\rangle \otimes |0...0\rangle) = (D_{2^{n}}|\psi\rangle)\otimes|0...0\rangle
\end{equation}

%\subsection{Circuit Cutting}

%Circuit cutting is a method for decomposing larger circuits into smaller subcircuits, with the trade-off of additional sampling overhead\cite{qiskit-addon-cutting}. Circuit width, depth, and swaps can be decreased to execute the ISA on hardware to improve performance. Each of the \(P\) subdomain encoded circuits is decomposed into a \(Q\) number of cut subcircuit. The results of these smaller circuits can be recombined to yield results equivalent to those of the original circuit, albeit with ideally higher fidelity.

\begin{comment}
\subsection{Transpiler}
Transpilation rewrites a quantum circuit to match the instruction set architecture (ISA) and topology of executable hardware. Gates are first decomposed into one and 2-qubit operations. Virtual qubits, which we interact with using the native programming language, are then unrolled and laid out onto physical qubits in a compatible one-to-one manner. The one and two-qubit gates are then translated into the hardware's particular ISA.

The previous stages of the transpiler tend to increase the gate count and depth of the quantum circuit. These increases can create propagation errors through the circuit. To mitigate this, the transpiler has an optimization step to eliminate or combine gates to yield identical quantum channels using an optimal SWAP mapping, an NP-Hard stochastic heuristic algorithm.
\end{comment}

\subsection{Two Level Decomposition}

\begin{figure}
    \centering
    \begin{subfigure}[b]{0.20\textwidth}
         \centering
         \includegraphics[width=\textwidth]{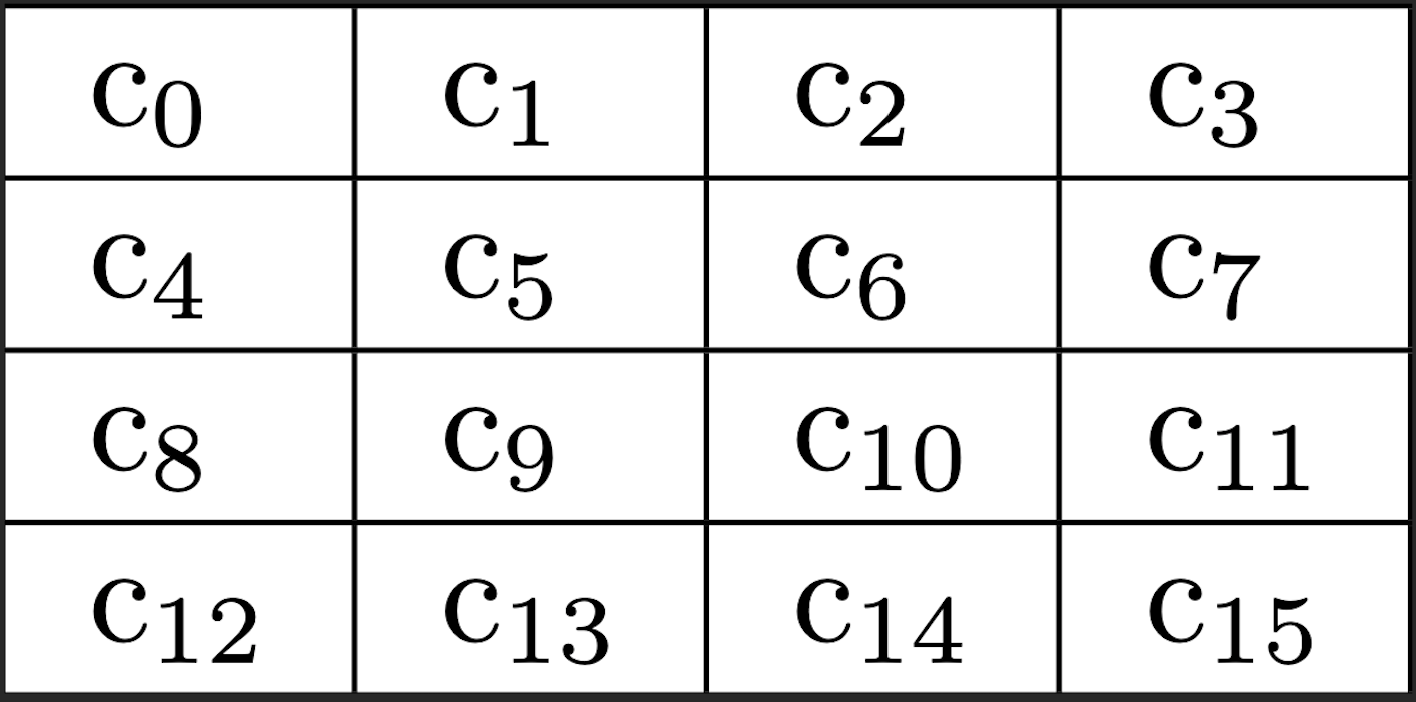}
     \end{subfigure}
     \begin{subfigure}[b]{0.35\textwidth}
         \centering
         \includegraphics[width=\textwidth]{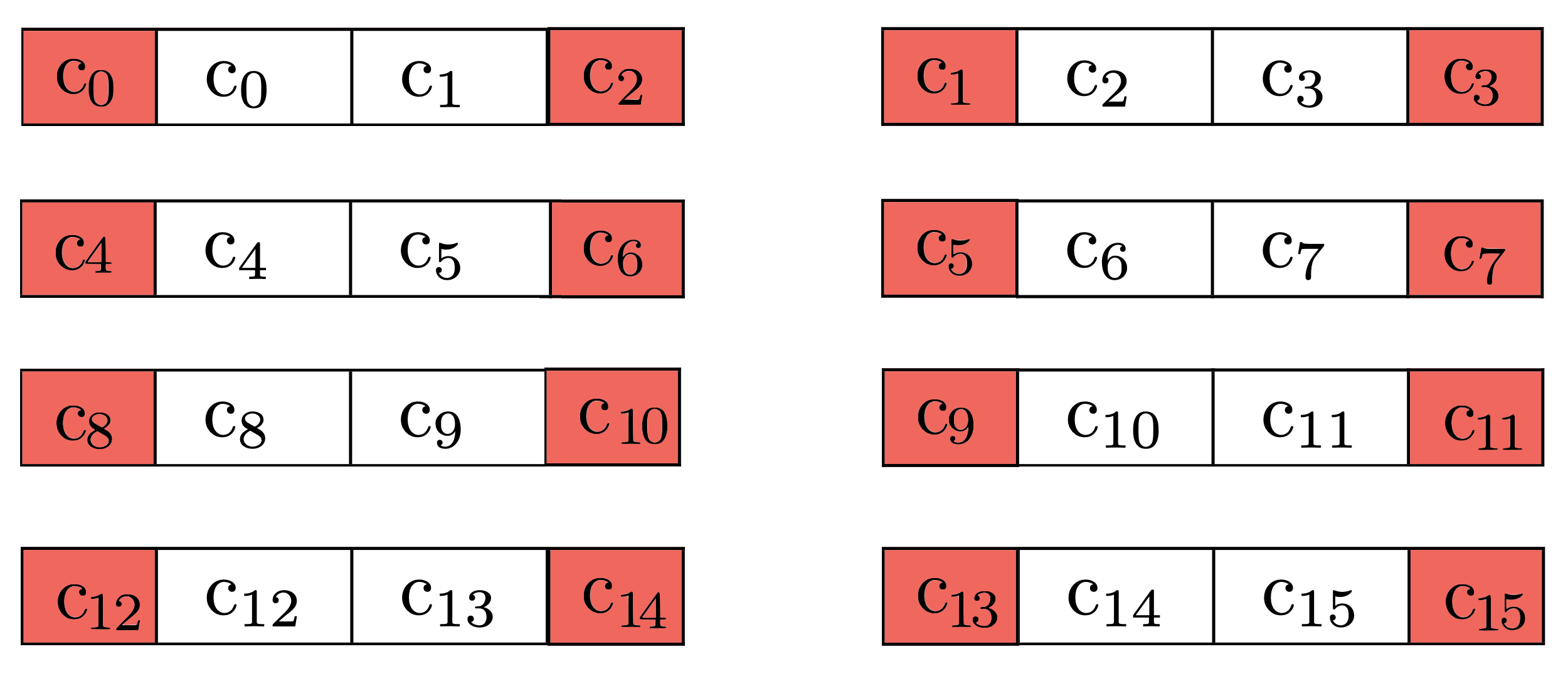}
     \end{subfigure}
    \caption{Mirrored buffer pixels are applied to the image boundaries. These neighboring pixels are included as a buffer for QHED error. Red cells are discarded in the final output.}
    \label{fig7}
\end{figure}

\begin{figure}
    \centering
    \includegraphics[width=0.95\linewidth]{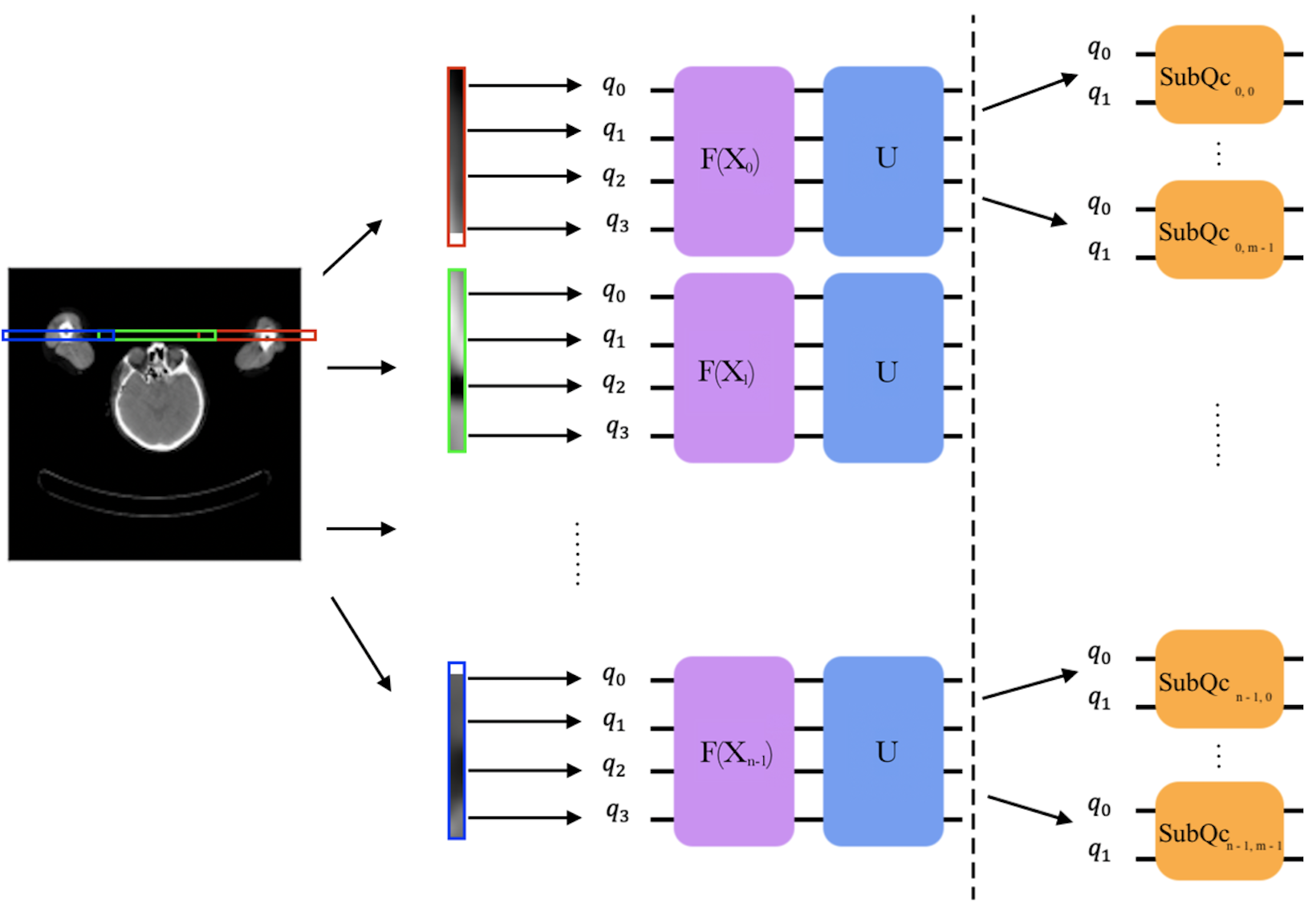}
    \caption{The D-NISQ methodology segments the input image into overlapping subdomains with buffer pixels to maintain boundary continuity. Utilizing a two-level decomposition approach, the resulting subcircuits are further partitioned to reduce circuit width, minimize depth, and decrease two-qubit operations, thereby improving fidelity. These optimized subcircuits are then distributed across a multi-node cluster for parallel processing. Finally, the results are reassembled to reconstruct an output equivalent to that of the non-decomposed statevector, ensuring computational accuracy while enhancing scalability.}
    \label{fig56}
\end{figure}

The two-level decomposition method consolidates image decomposition and circuit cutting in a two-step process. As shown in Figure \ref{fig56}, the input image is first divided into \(P\) subimages, with buffer pixel cells added between adjacent subdomains to ensure continuity and to mitigate false edges on the boundaries of sub-images. Regardless of the input vector size, this leaves two redundant pixels per sub-image as shown in Fig. \ref{fig7}. Each subdomain is then amplitude-encoded into a quantum circuit followed by the QHED\textsuperscript{M} permutation. Following this, all circuits undergo a fully optimized transpilation process to maximize efficiency and overall performance.

Each optimized subdomain circuit is further subdivided into smaller \(Q\) number of subdomains through circuit cutting. The number of qubits per subcircuit is limited to a predefined maximum, ensuring an optimal trade-off between computational efficiency and performance giving a total number of \(P \times Q\) circuits to be executed. These subcircuits are then distributed across a high-performance computing (HPC) cluster for independent evaluation. In the final step, the results of the subcircuits are recombined in reverse order to reconstruct the expected non-decomposed statevector.

\subsection{Quantum Resilience Layers}

To address the decoherence limitations identified in the circuit-level decomposition, we integrated a resilience layer focusing on both active and passive error management. The error management architecture presented in this study builds upon the foundational protocols established in the IBM Qiskit Global Summer School 2025 \cite{qgss2025-lectures}\cite{qgss2025-labs}. Specifically, the implementation of the $[\![3,1,1]\!]$ bit-flip repetition code and the comparative methodology for Estimator-based error mitigation were adapted from the QGSS 2025 curriculum on Quantum Error Correction. By leveraging the QGSS 2025 framework, we standardized our resilience testing environment. The approach to characterizing noise channels on superconducting backends provided the theoretical basis for our selection of the "Resilience Level 1" (readout error extinction) strategies used in the QEM trials. This pedagogical foundation ensures that our edge detection results are benchmarked against industry-standard correction protocols.

\subsubsection{Active Protection}

Given the dominance of relaxation ($T_1$) and dephasing ($T_2$) errors in superconducting transmon qubits, single-qubit bit-flip errors ($\sigma_x$) represent a significant failure mode during deep circuit execution. To actively suppress these errors, we employed a $[\![3,1,1]\!]$ repetition code. This scheme encodes a single logical qubit state $|\psi\rangle_L$ into a subspace of three physical qubits. The logical basis states are defined as entangled Greenberger–Horne–Zeilinger (GHZ)-like states: $$|\bar{0}\rangle \equiv |000\rangle, \quad |\bar{1}\rangle \equiv |111\rangle$$
The encoding unitary, $U_{enc}$, maps an arbitrary input state $|\psi\rangle = \alpha|0\rangle + \beta|1\rangle$ to the logical subspace using two CNOT gates, where the data qubit acts as the control and two ancilla qubits act as targets: $$|\psi\rangle_L = U_{enc} |\psi\rangle \otimes |00\rangle = \alpha|000\rangle + \beta|111\rangle$$
In our implementation, the Logical QHED operator $U_{QHED}$ is applied directly to the physical register. If a bit-flip error occurs on the $i$-th qubit during execution, the state typically exits the code space. For example, a flip on the first qubit results in $\alpha|100\rangle + \beta|011\rangle$. To recover the state, we apply a unitary decoding $U_{dec} = U_{enc}^\dagger$. In a standard error correction cycle, syndrome measurements would identify the error location. However, to minimize circuit depth on the FakeNairobiV2 backend, we utilize a post-selection strategy on the ancilla qubits. We trace the ancillary system and reconstruct the density matrix $\rho_{logical}$ only from shots where the ancillaries measure $|00\rangle$, effectively projecting the system back into the code space:$$\rho_{logical} = \frac{\Pi_0 \rho_{total} \Pi_0}{\text{Tr}(\Pi_0 \rho_{total})}$$ where $\Pi_0 = I \otimes |00\rangle\langle00|$

\begin{figure}
    \centering
    \includegraphics[width=1.0\linewidth]{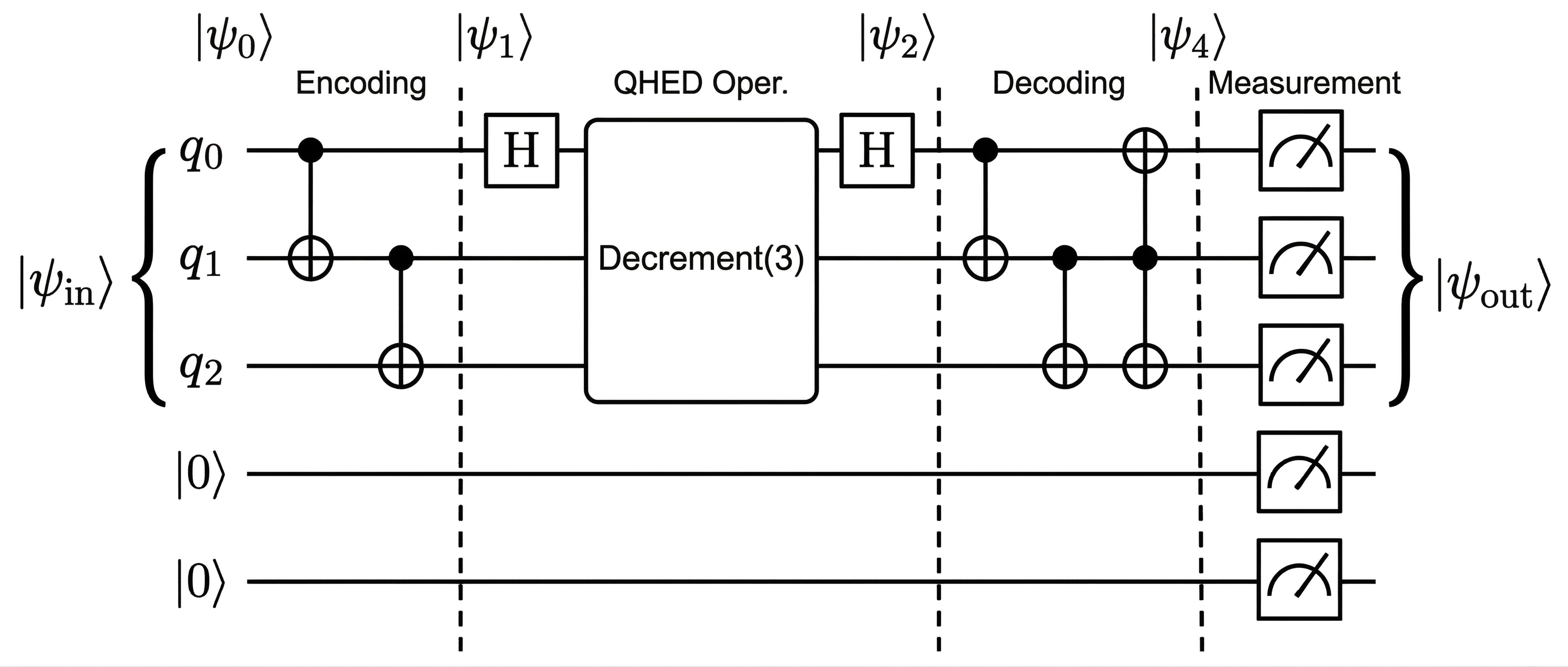}
    \caption{The logical QHED pipeline wrapped in a $[\![3,1,1]\!]$ repetition code. The logical operation is performed on the physical qubits between the encoding and decoding stages.}
    \label{fig:qc-diagram}
\end{figure}

\subsubsection{Passive Mitigation}

While active QEC changes the quantum state, Passive Quantum Error Mitigation (QEM) uses classical post-processing to refine the expectation values of observables. We utilize the Qiskit Runtime Estimator primitive with resilience level 1, which targets readout (measurement) errors. The readout errors are modeled by a classical assignment matrix $A$, where $A_{ij} = P(i|j)$ is the probability of observing the bit string $i$ given the state $j$. The noisy probability vector $\vec{P}_{noisy}$ is related to the ideal vector $\vec{P}_{ideal}$ by:$$\vec{P}_{noisy} = A \vec{P}_{ideal}$$
To recover the utility of QHED edge detection, we estimate the expectation value of the dominant edge intensity operator $O = Z \otimes I^{\otimes (n-1)}$. The mitigated expectation value is computed by applying the inverse of the calibration matrix:$$\langle O \rangle_{mit} = \sum_{x \in \{0,1\}^n} (\vec{P}_{noisy}^T A^{-1})_x \cdot \lambda_x$$ where $\lambda_x$ are the eigenvalues of $O$ associated with the basis state $x$.
\section{Evaluation and Results}

\begin{figure*}
    \centering
    \begin{subfigure}[b]{0.32\textwidth}
         \centering
         \includegraphics[width=\textwidth]{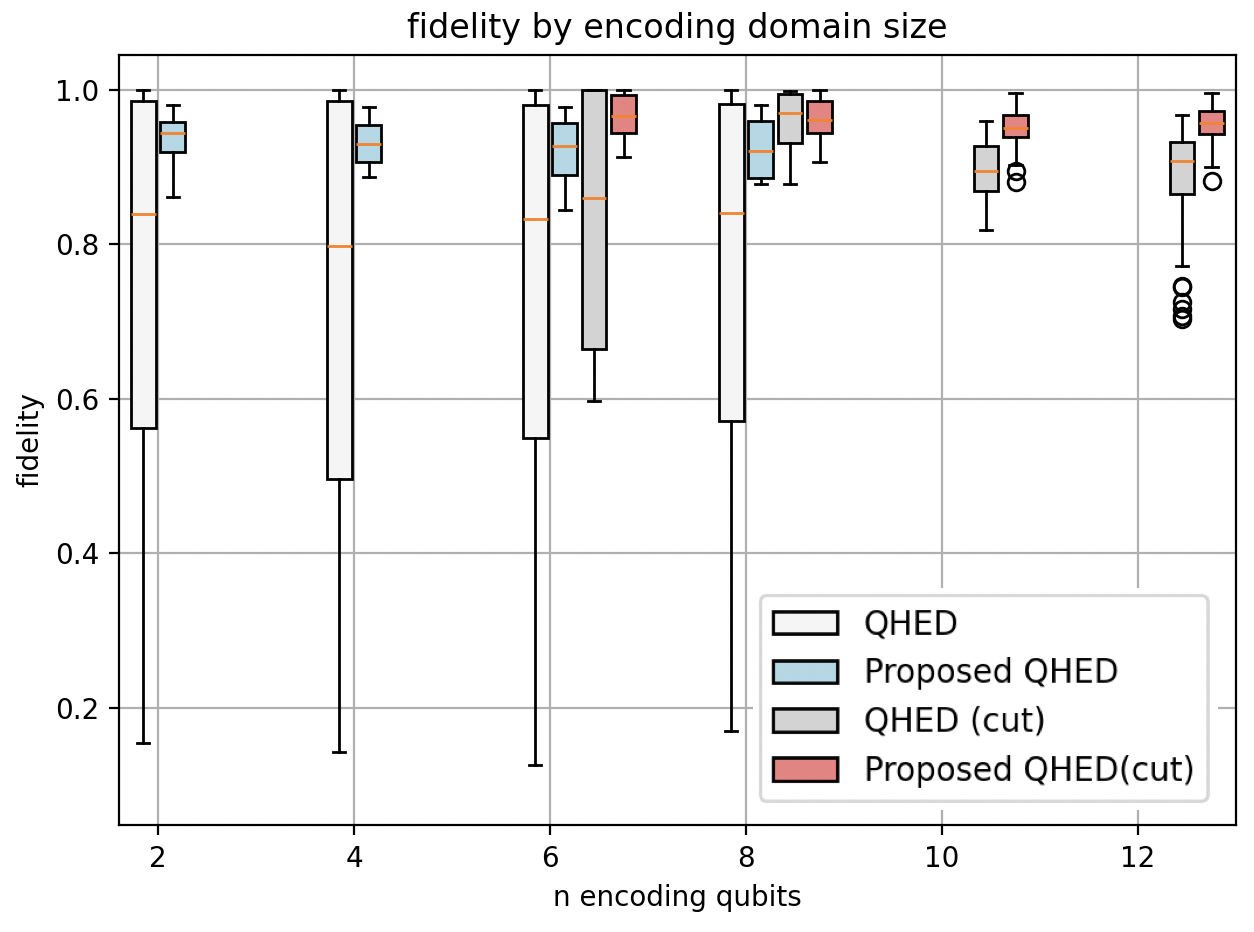}
         \caption{}
         \label{97a}
     \end{subfigure}
    \begin{subfigure}[b]{0.32\textwidth}
         \centering
         \includegraphics[width=\textwidth]{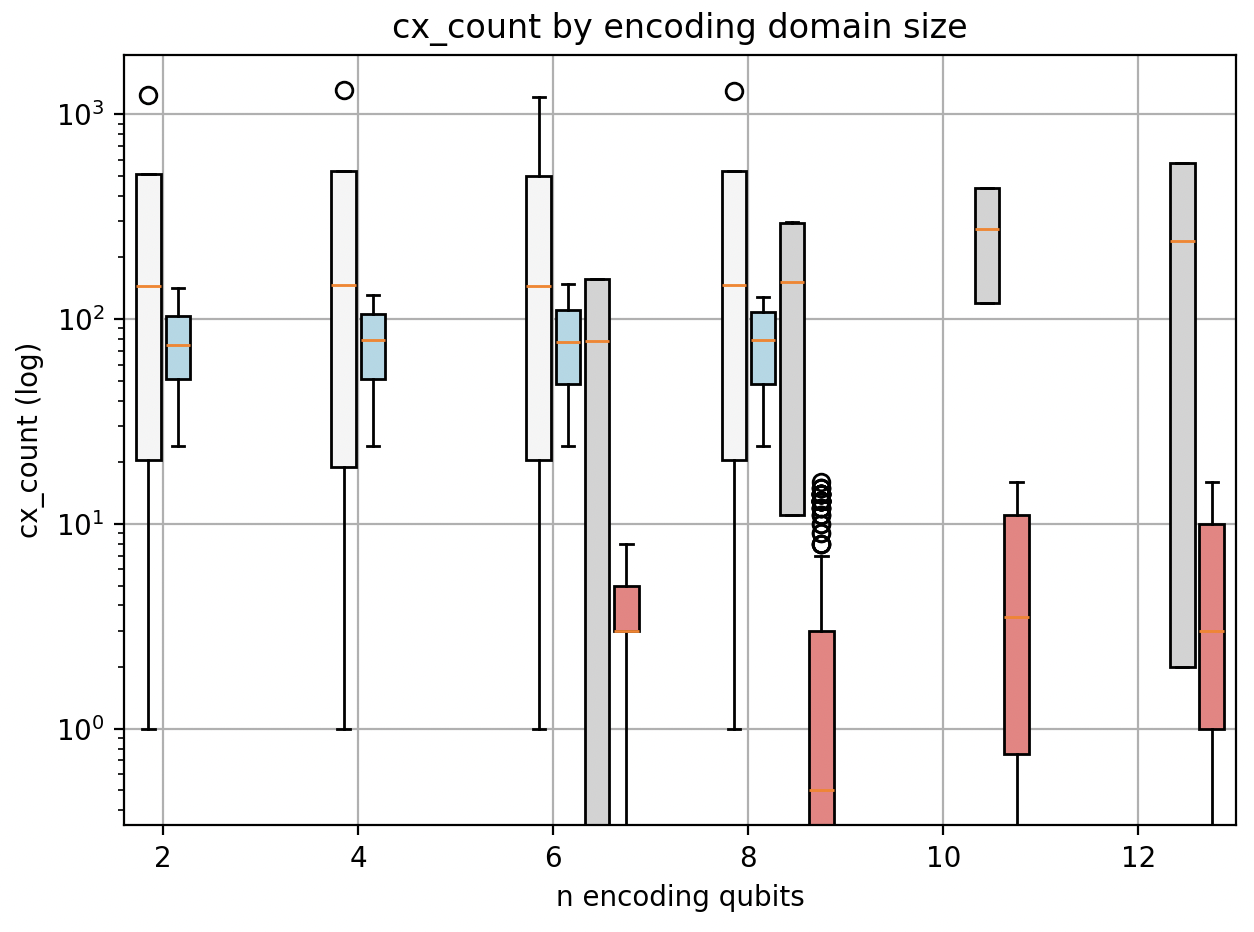}
         \caption{}
         \label{97b}
     \end{subfigure}
     \begin{subfigure}[b]{0.32\textwidth}
         \centering
         \includegraphics[width=\textwidth]{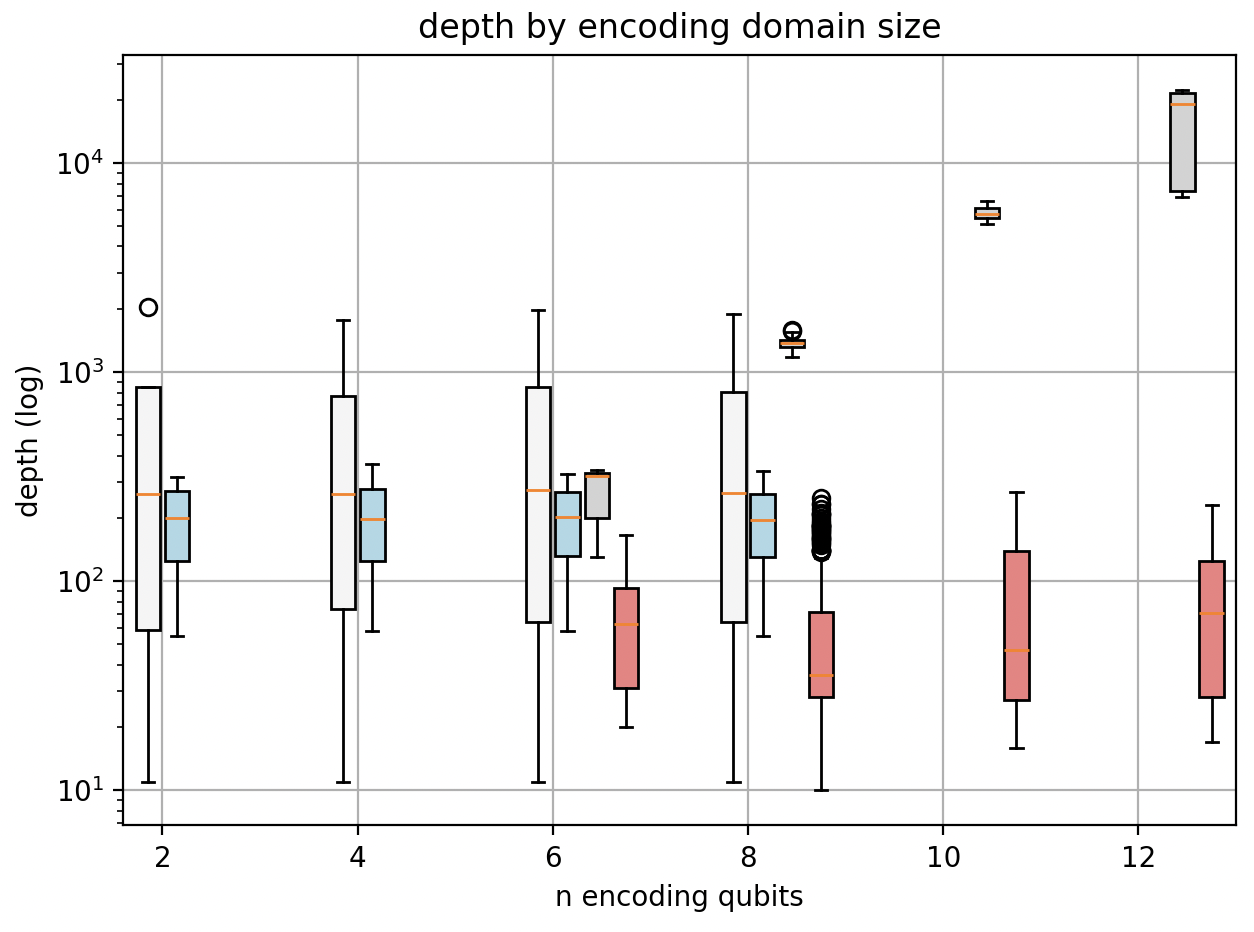}
         \caption{}
         \label{97c}
     \end{subfigure}
    \caption{
    Analysis of the original and proposed QHED algorithms using IBM's \(FakeMumbai\) backend, with full circuit optimization applied. Each circuit was executed 100 times—both as a whole and in its cut form—using different seeds for transpilation and execution. Original circuits were tested over encoding domain sizes ranging from 2 to 8 qubits, while cut circuits were evaluated from 6 to 12 qubits. (b) and (c) use a logarithmically scaled y-axis.}
    \label{97}
\end{figure*}

\begin{figure*}
    \centering
     \begin{subfigure}[b]{0.34\textwidth}
         \centering
         \includegraphics[width=\textwidth]{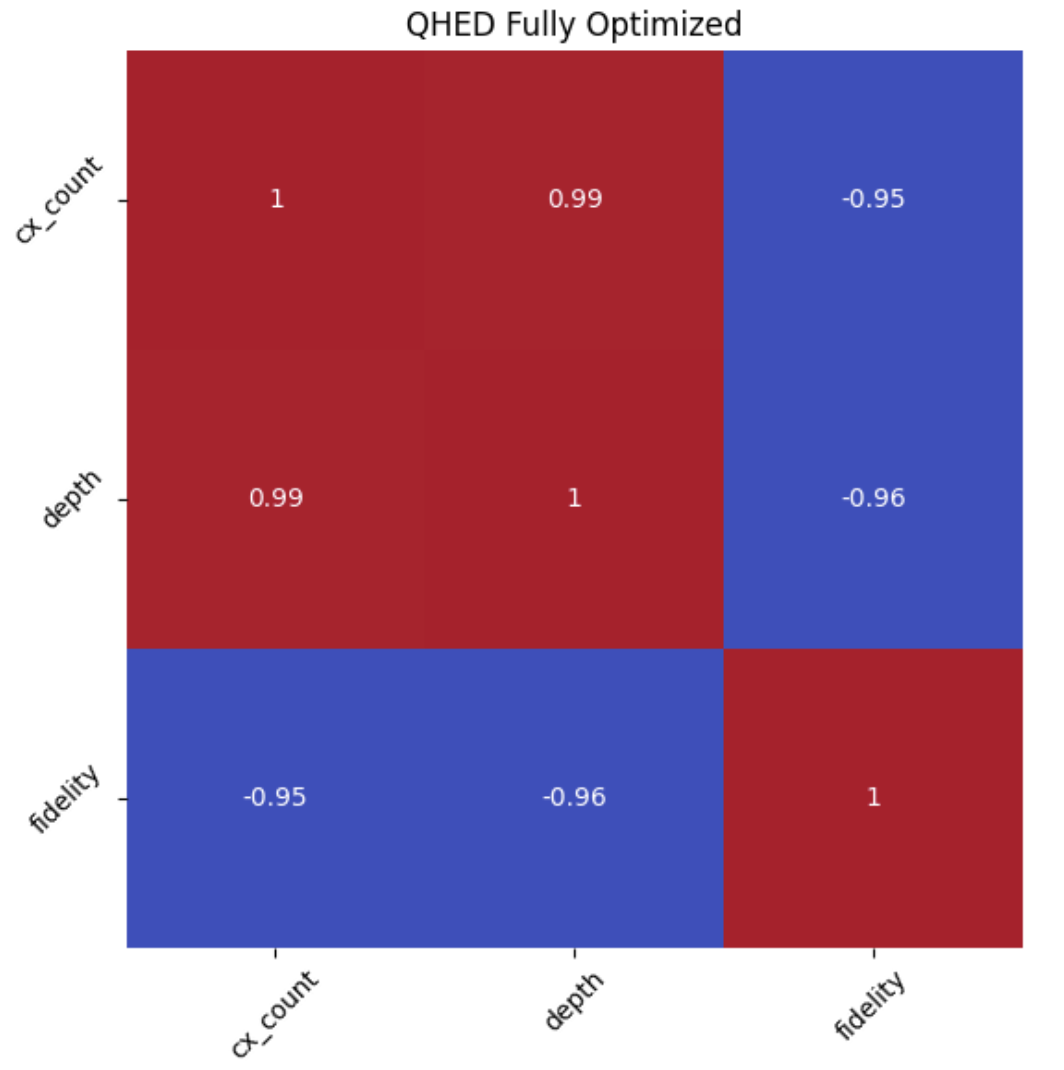}
         \caption{}
         \label{fig99a}
     \end{subfigure}
     \begin{subfigure}[b]{0.39\textwidth}
         \centering
         \includegraphics[width=\textwidth]{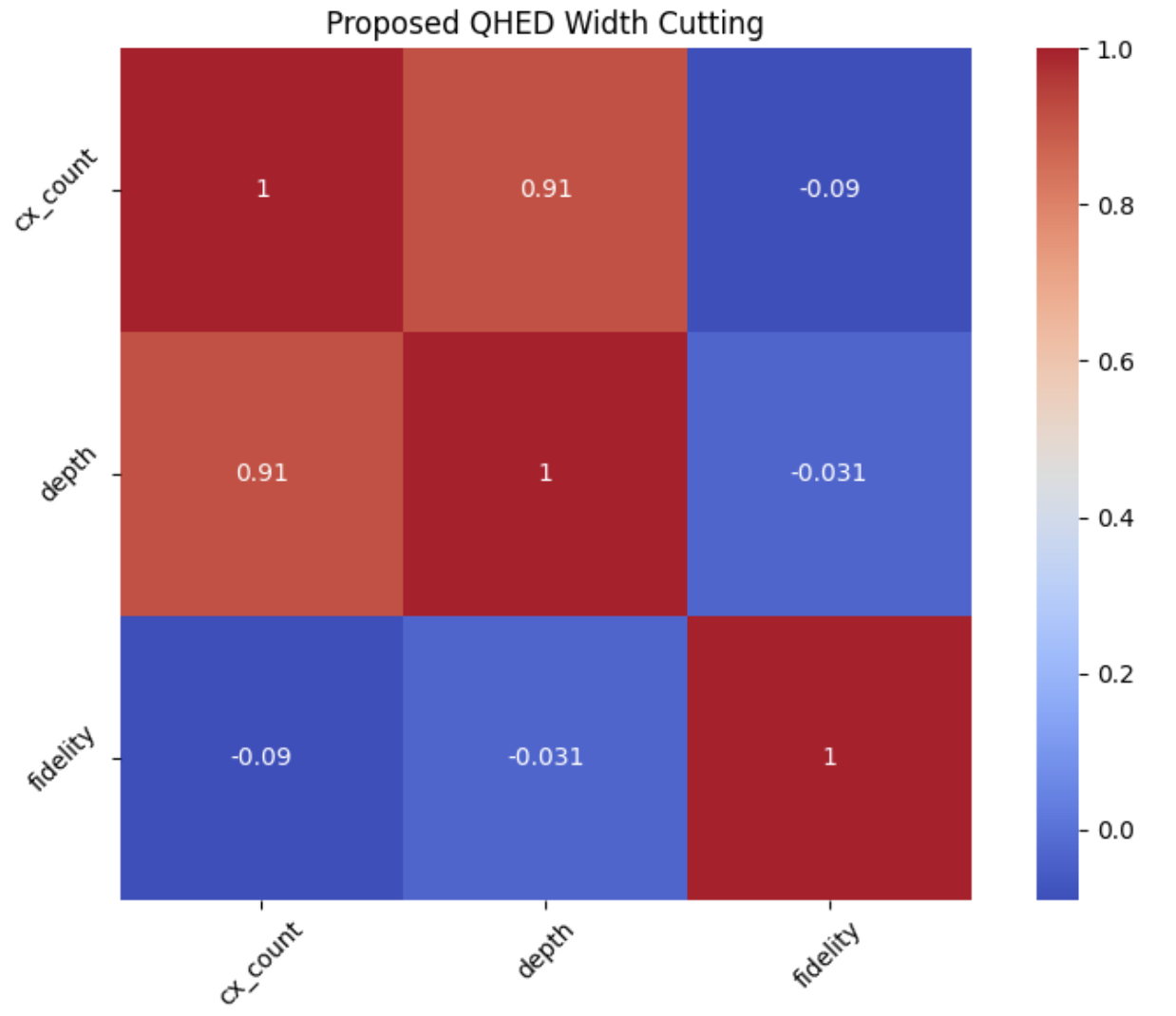}
         \caption{}
         \label{fig99b}
     \end{subfigure}
    \caption{Correlation matrices of our given metrics for both the original QHED circuit fully optimized (a) and the QHED\textsuperscript{M} circuit fully optimized with applied width cutting (b).}
    \label{fig99}
\end{figure*}

\subsection{Experimental Setup}

Our evaluation is conducted in two distinct stages. In the first stage, we perform a comprehensive analysis of the original QHED circuit alongside our proposed circuit, both with and without the application of circuit cutting techniques. At this stage, circuits are treated as abstract computational units capable of processing \(2^{n}\) input size on \(N\) qubit device, without encoding any actual data. Given that Yao et al did not produce a fidelity study in the original QHED publication, this abstraction allows us to establish a baseline for performance characteristics of each circuit architecture as a function of input size. We assess and compare the circuits based on fidelity, circuit depth, and CNOT (CX) gate count.

In the second stage, we apply these circuits to three distinct D-NISQ applications.

\begin{itemize}
    \item The first involves an experimental pipeline that processes raw k-space data as input. Using an ideal quantum simulator, we demonstrate the feasibility of combining the IQFT and the QHED\textsuperscript{M} circuit within a single pipeline to generate edge-detected images from real MRI k-space data. 
    \item The second application employs our QHED\textsuperscript{M} circuit on a 2D medical image with a Qiskit \(FakeMumbai\) noise model\cite{Qiskit}. Here, we demonstrate that our proposed image-level decomposition method, as illustrated in Fig.\ref{fig56}, enables the execution of this task on NISQ hardware in a distributed manner. We exclude circuit cutting due to current software limitations concerning the reassembling of bitstrings\cite{qiskit-addon-cutting}. 
    \item The third application extends the image-level decomposition approach to a 3D MRI dataset from the National Cancer Institute\cite{cancer_data_commons}, utilizing an ideal quantum simulator with the QHED\textsuperscript{M} circuit further to demonstrate the utility and scalability of the proposed methodology.
\end{itemize}

The D-NISQ model simulations were carried out on the Old Dominion University Wahab HPC cluster (2-10 nodes with Intel(R) Xeon(R) Gold 6148 2.4GHz CPUs, managed by SLURM\cite{10.1007/10968987_3}), leveraging the Dask Python framework\cite{Dask} for parallel execution (4096 shots per subcircuit). The proposed QHED optimizations were evaluated using a simulated noisy backend ('Fake Providers') from IBM Qiskit called \(FakeMumbai\)\cite{qiskit2024}\cite{Qiskit-Textbook}. Each circuit underwent thorough optimization and was executed with 100 different seeds for transpilation and execution across increasing encoding domain sizes \(n\), as illustrated in Fig. \ref{97}.

\subsection{Metrics}

Optimizing the design of an existing quantum circuit through targeted modifications and software-based optimization techniques has proven effective in enhancing performance on actual quantum hardware. This process simulated the fidelity of the first-stage (image) decomposition results with and without circuit cutting.

The width of each fully optimized circuit was limited to a maximum of five qubits per sub-circuit to balance computational overhead and minimize fidelity loss \cite{Chrisochoides_Liu_Drakopoulos_Kot_Foteinos_Tsolakis_Billias_Clatz_Ayache_Fedorov_et_al._2023a}. As the encoding domain size increases, the original circuit exhibits a broader distribution of results and a decline in average fidelity, as shown in Fig. \ref{97a} with the absence of circuit cutting. The wider spread of fidelity outcomes in Fig. \ref{97a} suggests that the standard QHED circuit yields inconsistent results on NISQ hardware. In contrast, our proposed QHED\textsuperscript{M} circuit demonstrates a tighter distribution with an overall higher average, suggesting significantly greater predictability and performance.

Fig. \ref{97b} highlights the substantial number of CX gates generated by the original QHED circuit. Once again, the original circuit exhibits inconsistencies, with variations that can be greater than those generated by our redesigned circuit. These inconsistencies persist in the cut circuits, whereas our QHED\textsuperscript{M} circuit significantly reduces the number of CX gates. Furthermore, as shown in Fig. \ref{97c}, our proposed circuit achieves a smaller and more consistent depth distribution. In contrast, even with circuit cutting, the original circuit exhibits an exponential increase in depth.

Encoding a 256x256x130 voxel MRI scan using the original QHED circuit would require 24 total qubits. However, based on the fidelity results presented in Fig. \ref{97}, this would output incomprehensible noise. In contrast, our two-level decomposition strategy, which partitions the volume into five-qubit subdomains (including buffer voxels), results in \(P=283,990\) subdomains. Using the original QHED circuit, each subdomain would require 6 qubits, whereas our QHED\textsuperscript{M} circuit requires 10 qubits per subdomain for increased fidelity.

\subsection{D-NISQ Application}

\begin{figure}
    \centering
    \includegraphics[width=0.95\linewidth]{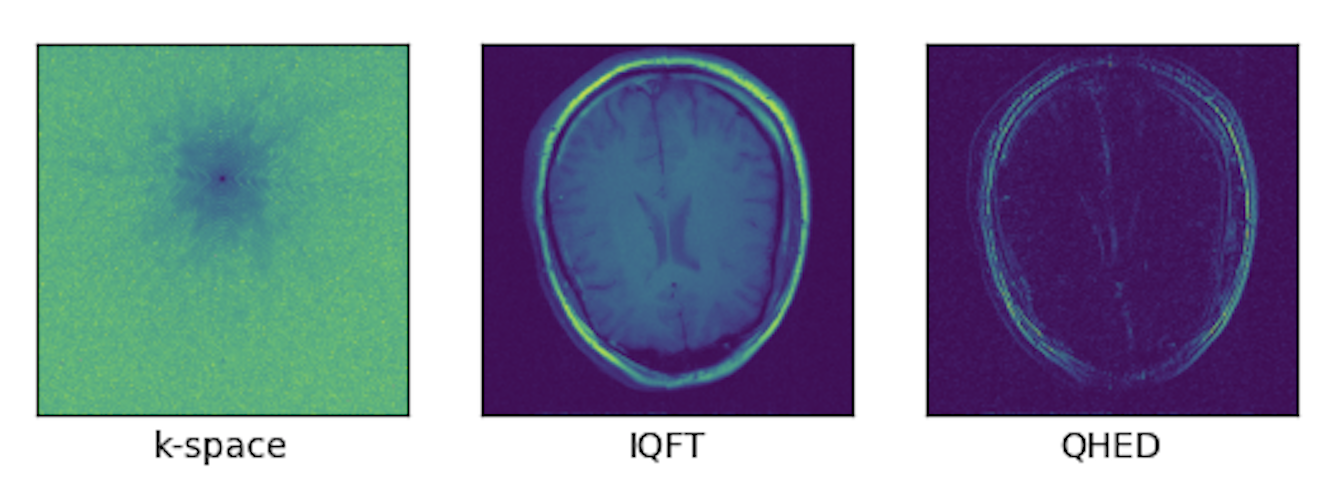}
    \caption{The K-space MRI output is normalized and processed through the IQFT pipeline using an ideal quantum simulator. Each stage of the sequence is normalized to real values for visualization purposes.}
    \label{fig98}
\end{figure}

The IQFT/QHED pipeline, shown in Fig. \ref{fig11}, is demonstrated using an ideal quantum simulator. 1D K-Space MRI data, 1024x1024 pixels, is amplitude encoded and subject to the IQFT circuit. The complete measurement of this output yields the normalized 1D medical image to be passed through the QHED\textsuperscript{M} circuit. The results are shown in Fig. \ref{fig98}.

\begin{figure}
    \includegraphics[width=0.45\textwidth]{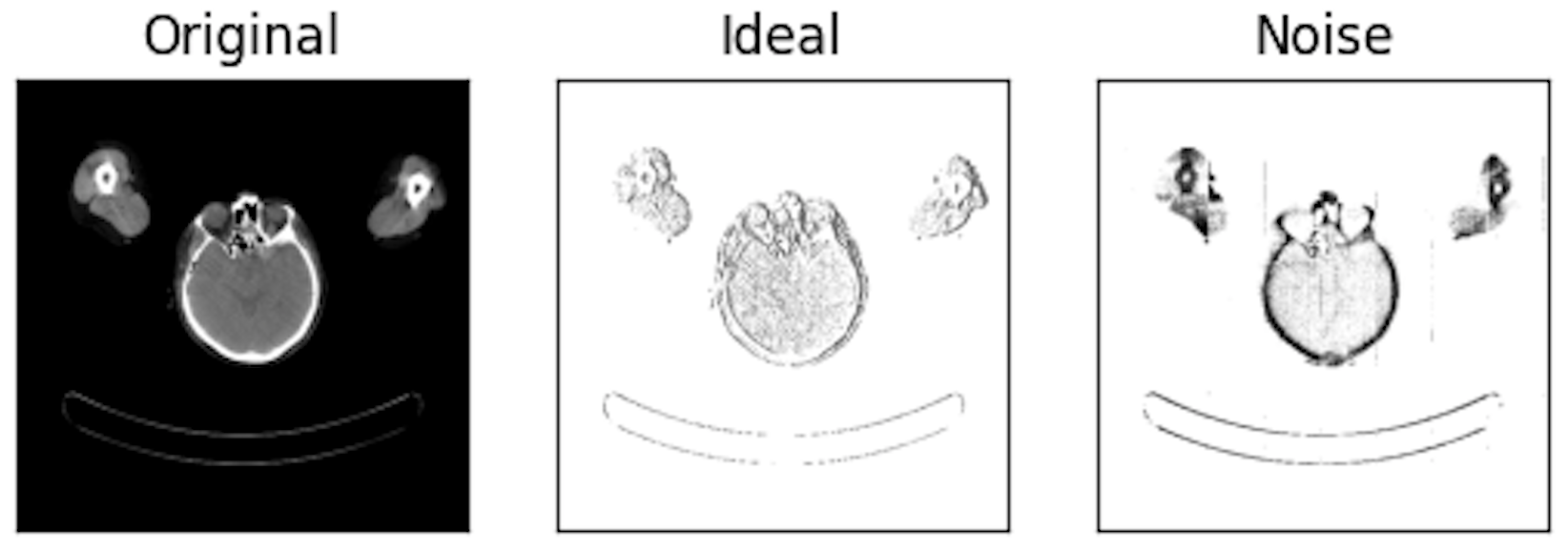}
    \caption{An image obtained from the National Cancer Institute is processed and passed through the QHED\textsuperscript{M} algorithm using both ideal and noisy quantum simulators. The images are generated using the image-level decomposition methodology on a dynamically scalable multi-node HPC cluster.}
    \label{fig12}
\end{figure}

The results of using a noisy simulation with applied optimization techniques of the QHED\textsuperscript{M} circuit at the image decomposition level are shown in Fig. \ref{fig12}. The 1024x1024 image is buffered with its domains decomposed to 5 qubits (or 32 pixels). The Dask scheduler, which is responsible for resource management and execution of computational tasks, distributes each subdomain to a dynamically scaled (between 2 and 10 nodes based on resource availability) cluster, which acts as our simulated QPU's in the fashion of a D-NISQ simulation. These domains are encoded into the QHED\textsuperscript{M} circuit and executed on an ideal simulator as well as a noisy simulator. The results are then reassembled with threshold values applied for visual purposes.

\begin{figure}
    \centering
    \begin{subfigure}[b]{0.48\textwidth}
    \includegraphics[width=\textwidth]{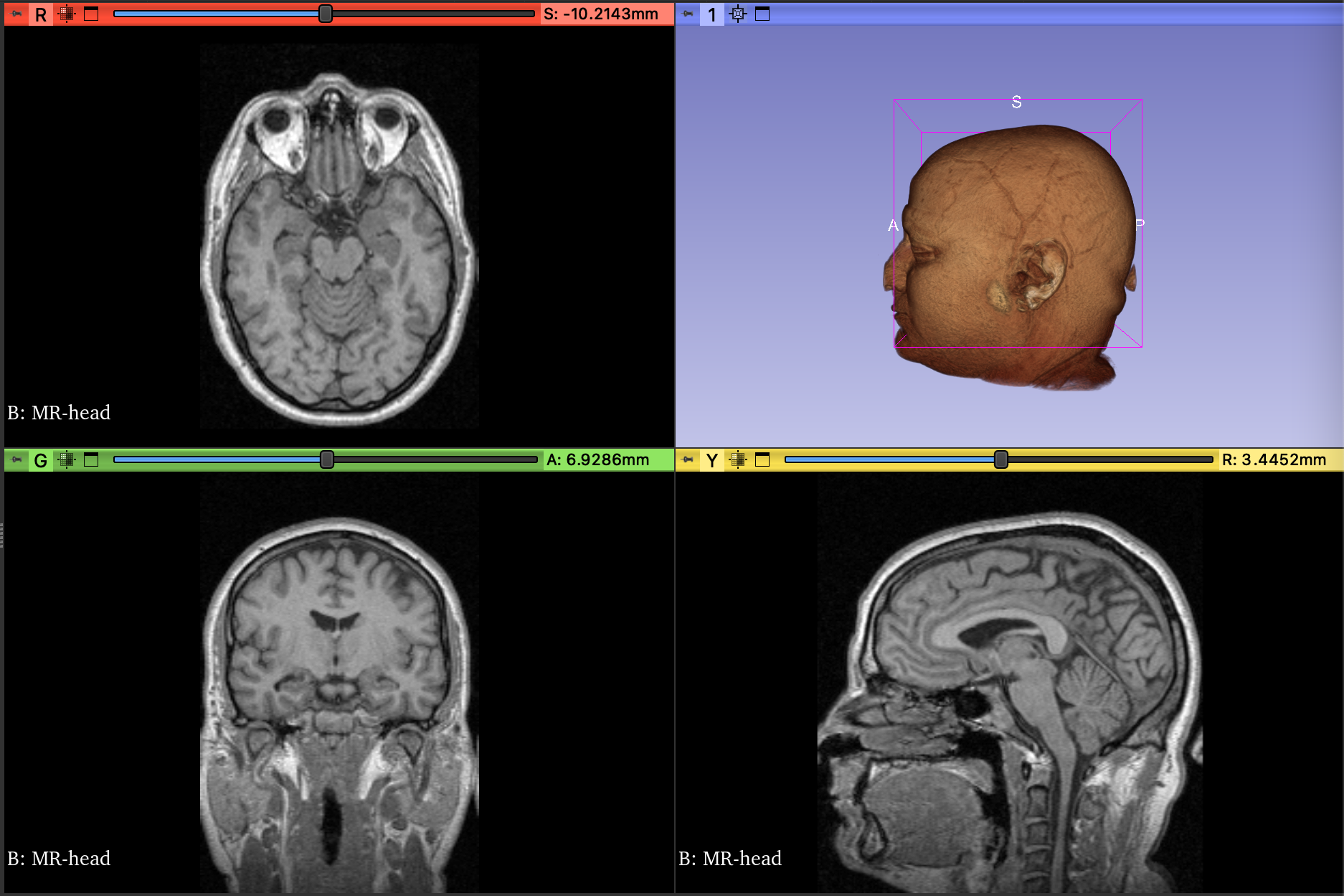}
    \end{subfigure}
    \begin{subfigure}[b]{0.48\textwidth}
    \includegraphics[width=\textwidth]{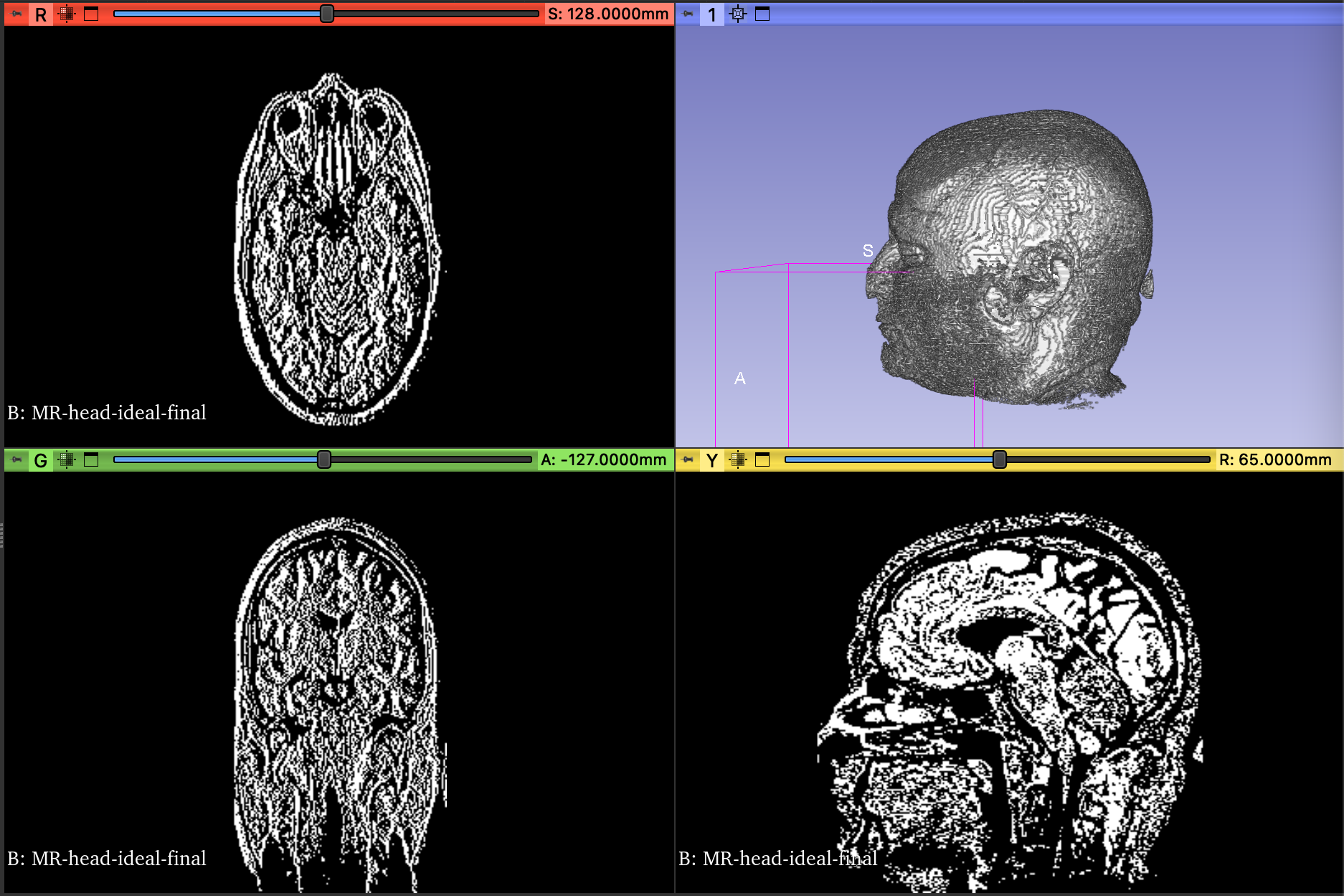}
    \end{subfigure}
    \caption{The 3D MRI model, sourced from the National Cancer Institute (top), is processed and passed through the distributed QHED\textsuperscript{M} algorithm using an ideal quantum simulator (bottom). The images are generated using the image-level decomposition methodology on an HPC cluster.}
    \label{fig13}
\end{figure}

\begin{figure}
    \includegraphics[width=0.45\textwidth]{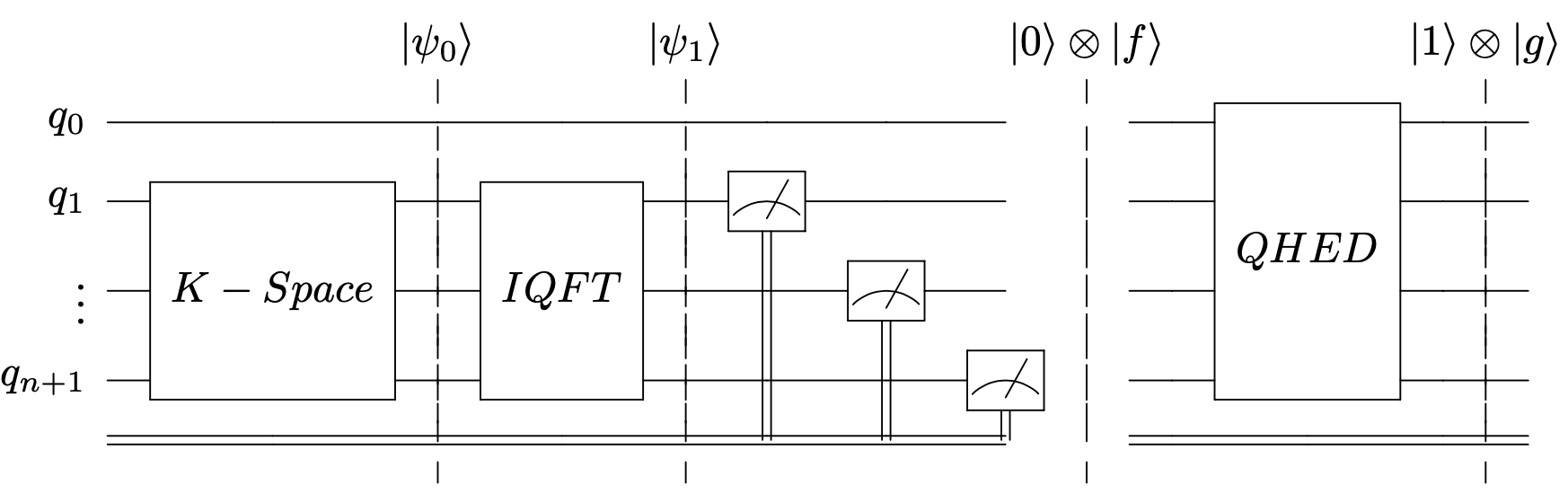}
    \caption{The circuit pipeline necessary for implementing the QHED algorithm to perform edge detection on MRI k-space data. This pipeline includes various stages of data processing, from initial encoding of the MRI k-space information into quantum circuits, to the application of the QHED algorithm for feature extraction. The process is designed to enhance the resolution of edges within the MRI data, leveraging quantum optimization techniques for improved accuracy and performance.}
    \label{fig11}
\end{figure}

Fig. \ref{fig13} shows the result of the QHED algorithm applied to a 3D model. This result was gained using an ideal simulator at the image decomposition level with a D-NISQ simulation. The model size is 256x256x130 voxels which is then buffered and decomposed into 5 qubit (32 voxel) subdomains to be distributed. The output of each subdomain is collected, post-processed, and reassembled. Threshold values are applied for visual purposes with the output shown in Fig. \ref{fig13}.

\subsection{Resilience Analysis \& Utility Assessment}

We evaluated the efficacy of the resilience layers by comparing the state fidelity and expectation values of the QHED circuit under three conditions: unprotected noisy baseline, active QEC protection, and passive QEM. Experiments were conducted on the FakeNairobiV2 backend with 8192 shots.

\subsubsection{The Cost of Active Correction}

Contrary to ideal theoretical predictions, the application of the $[\![3,1,1]\!]$ code resulted in a net decrease in fidelity compared to the unprotected baseline. The baseline noisy fidelity was 0.9553, whereas the QEC-protected fidelity dropped to 0.3945. This degradation highlights gate overhead penalty as a critical NISQ-era constraint. The encoding and decoding stages introduce $4$ additional CNOT gates per logical qubit. On the FakeNairobi topology, which has limited connectivity, mapping these CNOTs requires additional SWAP gates. The cumulative error probability introduced by these extra gates ($p_{gate} \times N_{added}$) exceeded the error probability of the bit-flips they were intended to correct ($p_{flip} \times N_{original}$).$$\epsilon_{QEC} > \epsilon_{Baseline} \implies \text{Overhead} > \text{Correction Gain}$$

\subsubsection{The Utility of Passive Mitigation}

\begin{figure}
    \centering
    \includegraphics[width=1.0\linewidth]{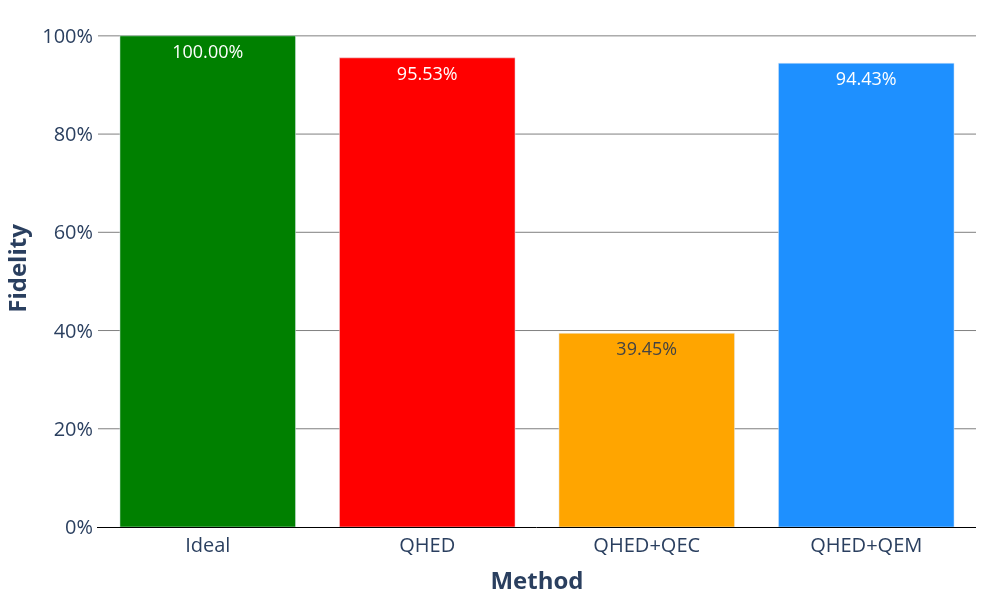}
    \caption{Comparison of edge detection signal intensity (Expectation Value of ZII). Passive QEM (Level 1) recovers nearly 99\% of the ideal signal, whereas active QEC degrades performance due to excessive gate overhead.}
    \label{fig:expectated-value-comparison-chart}
\end{figure}

Passive QEM yielded significantly higher utility. By bypassing the gate overhead and focusing solely on correcting the measurement distribution, the estimator (resilience level 1) recovered the expectation value of the edge operator $\langle ZII \rangle$. As illustrated in Fig. \ref{fig:expectated-value-comparison-chart}, the ideal expectation value for the target edge state was 1.0. The noisy baseline deviated to 0.9553. Active QEC, due to the scrambling effects of the overhead, drifted further to 0.3945. However, passive QEM successfully recovered the value to 0.9443, yielding a relative error of only 1.15\% compared to the ideal.

\section{Cross-Domain Generalization}

While the primary validation of our decomposition strategy focused on k-space MRI data, the mathematical abstraction of Quantum Hadamard Edge Detection (QHED) is inherently domain-agnostic. The algorithm operates not on the semantic meaning of the data (e.g., biological tissue vs. asphalt roads), but on the probability amplitudes of the encoded quantum state. In this section, we extend our framework to Geospatial Information Systems (GIS) and Remote Sensing, demonstrating utility in coastline monitoring, disaster response (wildfires), and precision agriculture.

\subsection{The Abstraction of Amplitude Encoding}

The core advantage of our architecture lies in the mapping of classical information vectors $\vec{x}$ to quantum amplitudes. Whether $\vec{x}$ represents the proton density of a brain scan, the thermal infrared signature of a forest fire, or the Near-Infrared (NIR) reflectance of a crop field, the encoding remains:$$|\psi_{data}\rangle = \sum_{i=0}^{N-1} x_i |i\rangle$$
In the context of Remote Sensing, high-resolution satellite imagery (e.g. Sentinel-2, Landsat, or MODIS) presents a massive data volume challenge similar to MRI, but at a planetary scale. A standard 100km $\times$ 100km satellite tile may contain over $10^8$ pixels. Classical edge detection algorithms scale linearly $O(N)$ with the number of pixels. In contrast, our QHED approach scales logarithmically $O(\log N)$ regarding the number of qubits required to represent the indices.

\subsection{Application 1: Coastal Erosion Monitoring}

We first applied the $P \times Q$ decomposition strategy to synthetic datasets representative of high-contrast coastal topology. In this abstraction, the binary transition between water (low reflectance) and land (high reflectance) creates a sharp gradient. The QHED circuit computes the discrete difference: $$|\psi_{out}\rangle \propto \sum_{x,y} (I(x,y) - I(x-1, y)) |x,y\rangle$$
By utilizing the Passive QEM strategies detailed in Section IV, we can extract coastline vectors from noisy quantum measurements with high fidelity, enabling automated change detection for rising sea levels without the computational bottleneck of classical pixel-by-pixel convolution.

\subsection{Application 2: Wildfire Tracking}

A critical application for real-time edge detection is the identification of active wildfire fronts using thermal infrared telemetry. Unlike static coastlines, wildfire fronts are dynamic and characterized by extreme gradients in the Thermal emission bands (e.g., 3.9 $\mu m$).

\begin{itemize}
    \item \textit{Encoding}: We map the thermal intensity values $T(x,y)$ to the quantum amplitudes.
    \item \textit{Detection}: The Decrement Gate $U_{dec}$ acts as a high-pass filter. In a thermal image, the "edge" corresponds to the active fire line where the temperature spikes drastically against the background vegetation.
    \item \textit{Utility}: Rapidly identifying this vector boundary allows for automated resource allocation. The logarithmic scaling of QHED is particularly vital here, as it theoretically allows for the processing of hyper-spectral data cubes from geostationary satellites (such as GOES-R \cite{geos-r}) in near real-time, aiding in immediate disaster response.
\end{itemize}

\begin{figure}
    \centering
    \includegraphics[width=1.0\linewidth]{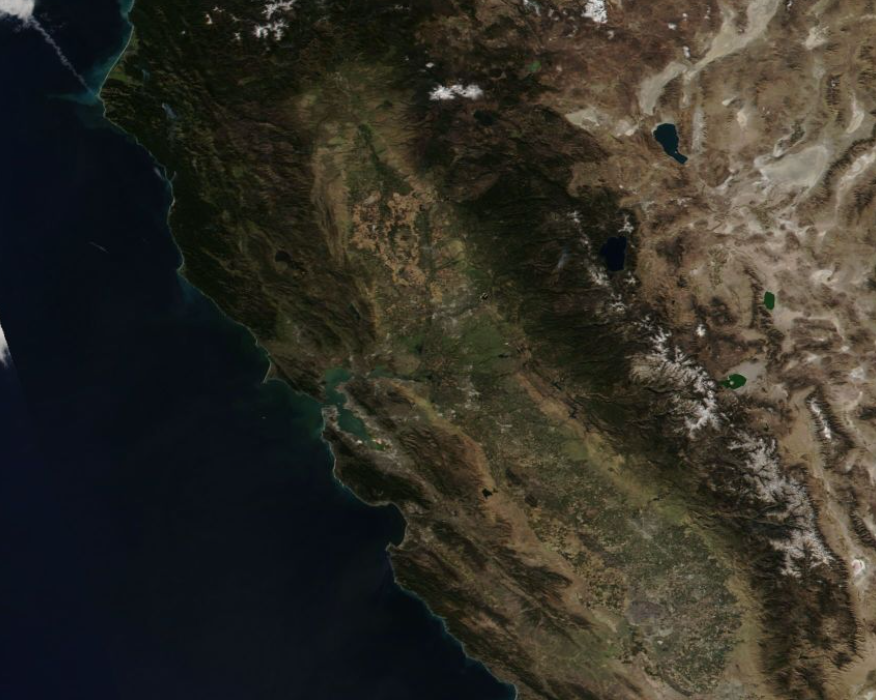}
    \caption{Geosationary satellite imagery of California for October 28, 2025. Imagery reflectance corrected is provided by the Visible Infrared Imaging Radiometer Suite (VIIRS) from the NASA/NOAA NOAA-21 (JPSS-2) satellite. Accessed through NOAA Worldview.}
    \label{fig:geos-california-10-2025}
\end{figure}

\subsection{Application 3: Agricultural Land Use \& Precision Farming}

In the domain of precision agriculture, defining field boundaries and crop heterogeneity is essential for yield estimation. We propose applying QHED to Normalized Difference Vegetation Index (NDVI) maps.

\begin{itemize}
    \item \textit{The Gradient}: Monoculture fields exhibit sharp boundaries (high gradients) between crops and service roads or fallow land. Within a field, "soft edges" indicate gradients in chlorophyll content, signaling pest infestation or irrigation leaks.
    \item \textit{Noise Resilience}: Agricultural imagery is often plagued by atmospheric scattering (noise). Our experimental results with Passive QEM (Section IV-C) demonstrate that the Estimator primitive effectively filters out uncorrelated noise while preserving the coherent gradient signal of the crop boundaries. This suggests that quantum edge detection could serve as a robust pre-processor for automated land-use classification pipelines.
\end{itemize}

\subsection{Scalability in Distributed Quantum GIS}

The data-level decomposition (splitting images into sub-patches) is particularly advantageous for these GIS applications. Unlike MRI data, which is locally bounded, satellite data is continuous. Our framework allows for a Distributed Quantum Computing (DQC) approach:

\begin{itemize}
    \item \textit{Tessellation}: A global map (coastal, forest, or farmland) is tessellated into $N$ tiles.
    \item \textit{Parallel Execution}: Each tile is processed as an independent quantum job.
    \item \textit{Reconstruction}: The edge-detected vectors are classically stitched back together.This implies that our findings on FakeNairobiV2 serve as a unit test for a massively parallelized quantum mapping engine capable of processing planetary-scale datasets.
\end{itemize}
\section{Conclusion and Future Work}

This work demonstrates that a restructured decrement permutation yields a substantial fidelity enhancement in the QHED circuit by reducing circuit depth from exponential to linear scaling with respect to input size as depicted in Fig. \ref{97c}. When integrated with circuit cutting techniques, these architectural optimizations yield further fidelity improvements, as depicted in Fig. \ref{fig99}, subject to an increased sampling overhead associated with the post-processing required for stitching subcircuit results. We have shown the feasibility of constructing a computational pipeline that begins with k-space data input reducing the computational complexity from \(O(N^{2})\) to \(O(N \ log N)\) and culminates in a distributed quantum edge detection algorithm, enabling image analysis and feature extraction with viable results as shown in Fig. \ref{fig98}. Although the image encoding of \(2^{n}\) amplitudes would require \(O(2^{n})\) or \(O(N)\) gates in the worst case, we aim for a future solution to the pipeline where only one encoding would be necessary.

Our approach has been validated for both two-dimensional (2D) pixel-based and three-dimensional (3D) voxel-based representations. Given real, positive-valued pixels and voxels, utility-scale quantum hardware can be leveraged for medical image processing. In future work, we aim to conduct a more thorough quantitative analysis of performance on 2D and 3D noise based models.

While our analysis of Fig. \ref{fig99} reveals that reducing the width of the proposed quantum circuit (QHED\textsuperscript{M}) achieves a significant decoupling of CX gate count and fidelity, this improvement currently incurs an exponential classical overhead due to the necessary knitting step following circuit cutting. To enhance the scalability and practical applicability of this approach, future research will focus on minimizing this classical overhead. We believe that continued efforts in optimizing the knitting process and exploring alternative circuit cutting and reconstruction techniques will make our approach increasingly more scalable, ultimately leveraging the noise resilience afforded by the decoupling of CX gates and fidelity.

% Added by David
% Our experiments revealed a critical crossover point for Utility-Scale Quantum Edge Detection. The implementation of the bit-flip code increased the circuit depth by a factor of $\approx 2.5\times$ due to the nearest-neighbor connectivity constraints of the Nairobi backend.While the QEC circuit successfully identified error syndromes, the decoherence accumulated during the syndrome extraction and decoding stages outweighed the correction benefits. In contrast, the QEM approach (Level 1) operated on the native decomposed circuit structure. By shifting the complexity from quantum gates (hardware) to classical post-processing (software), QEM maintained the structural integrity of the edge detection algorithm while filtering out readout noise. This suggests that for immediate real-world medical imaging applications, a hybrid pipeline of data-level decomposition + passive QEM is the optimal strategy.

Our experiments revealed a critical crossover point for Utility-Scale Quantum Edge Detection. The primary challenge in processing medical image data is not merely the qubit count, but the depth of the circuit required to encode complex gradient features. While the data-Level and circuit-Level decompositions proposed in Section III-A and III-B successfully fit the problem onto the chip, the Quantum Resilience Layer determines the quality of the output. The experimental data confirms that for current coherence times ($T_1 \approx 100 \mu s$), active error correction is premature for QHED workflows. The depth overhead of encoding destroys the quantum information faster than the code can protect it. Therefore, we conclude that the optimal pipeline for Utility-Scale QHED in the near term is a hybrid approach: data-Level decomposition to parallelize the image, coupled with passive Quantum Error Mitigation (QEM) to refine the gradients. This combination maximizes the "effective quantum volume" available for image processing without exceeding the hardware's decoherence threshold.
The versatility of the QHED operator across these domains—from detecting tumors in MRI to tracking wildfire fronts in thermal satellite data—confirms its role as a universal quantum signal processor. The commonality is the mathematical need to extract high-frequency spatial components (edges) from large, noisy datasets. Whether the noise comes from the MRI RF coil or atmospheric scattering in a satellite sensor, the combination of data-level decomposition and passive error mitigation provides a consistent pathway to utility.

Another promising direction for future research is the full distribution of the pipeline, starting from the inverse quantum Fourier transform (IQFT) circuit. Additionally, while it is theoretically possible to reconstruct the original state vector from cut circuit state vectors, current software limitations prevent direct implementation. A complete IQFT-QHED pipeline would involve the decomposition of k-space data into subdomains, encoding and partitioning each subdomain, distributing all circuit segments, and subsequently reassembling the computed state vectors. This process would ensure efficient quantum-based image processing by leveraging more advanced architectures \cite{Ang_Carini_Chen_Chuang_Demarco_Economou_Eickbusch_Faraon_Fu_Girvin_et_al._2024}.

The long-term objective of this study is to synergistically combine the ability of quantum sensing to gather more precise raw data (than current intraoperative MRI), thereby enhancing K-space data quality, with quantum computing's capacity to process intraoperative data rapidly. This could be a transformative leap in IGNS and general medical image computing \cite{Solenov2018}, \cite{National_Institutes_of_Health}, \cite{McWeeney_Perciano_Susut_Chatterjee_Fornari_Biven_Siwy_2023}.

\section*{Acknowledgements}

This research was supported in part by the Richard T. Cheng Endowment at Old Dominion University (ODU). The authors thank Min Dong at ITS in ODU for his help with HPC cluster at ODU. This work was performed using computational facilities at ODU enabled by grants from the National Science Foundation (MRI grant no CNS-1828593) and Virginia's Commonwealth Technology Research Fund. Any subjective views or opinions expressed in this paper do not necessarily represent the views of the National Science Foundation or the United States Government.

Gemini and Grammarly were used to improve readability; the authors reviewed and take full responsibility for the final content.

%\section*{Acknowledgment}
%This research was sponsored by the Richard T. Cheng Endowment. 

%\input{Sections/Appendix}

\bibliographystyle{IEEEtran}

\end{document}